\documentclass{ccjnl}

\usepackage{booktabs}
\usepackage{ragged2e}
\usepackage{array}
\usepackage{booktabs}
\usepackage{tikz}
\usepackage{IEEEtrantools}
\usepackage{multirow}
\usepackage{tabularx}
\usepackage{graphicx}
\usepackage{subcaption}
\usepackage[table]{xcolor}

\title{ Low-Altitude Wireless Networks: A Comprehensive Survey}

\author{Jun Wu\inst{1,+}, Yaoqi Yang\inst{2,+}, Weijie Yuan\inst{1,*}, Wenchao Liu\inst{1}, Jiacheng Wang\inst{3}, Tianqi Mao\inst{4},  Lin Zhou\inst{1},  Yuanhao Cui\inst{5}, Fan Liu\inst{6}, Geng Sun\inst{9}, Yiyan Ma\inst{11}, Nan Wu\inst{4,*}, Dezhi Zheng\inst{4}, Jindan Xu\inst{6}, Nan Ma\inst{5}, Zhiyong Feng\inst{5}, Wei Xu\inst{6}, Dusit Niyato\inst{3}, Chau Yuen\inst{3}, Xiaojun Jing\inst{5}, Zhiguo Shi\inst{7}, Bo Ai\inst{12,*}, Shi Jin\inst{6},  Dong In Kim\inst{8}, Jiangzhou Wang\inst{6}, Ping Zhang\inst{5}, Hao Yin\inst{10,*}, Jun Zhang\inst{4,*}\corinfo{zhjun@bit.edu.cn,  yinhao@cashq.ac.cn, boai@bjtu.edu.cn, wunan@bit.edu.cn, yuanwj@sustech.edu.cn}}
\receiveddate{Sep. 25, 2019} 
\reviseddate{Mar. 24, 2020}
\Editor{Wei Ma}
\address[1]{Southern University of Science and
Technology, Shenzhen 518055, China }
\address[2]{National Key Laboratory on Near-Surface Detection, Beijing 100072, China}
\address[3]{Nanyang Technological University, Singapore}
\address[4]{Beijing Institute of Technology, Beijing 100081, China} 
\address[5]{Beijing University of Posts and Telecommunications, Beijing 100876, China.}
\address[6]{Southeast University, Nanjing 210096, China.}
\address[7]{Zhejiang University, Hangzhou 310027, China.}
\address[8]{Sungkyunkwan University, Suwon 16419, South Korea. }
\address[9]{Jilin University, Changchun 130012, China.}
\address[10]{Institute of China Electronic System Engineering, Beijing 100141, China.}
 \address[11]{Beijing Jiaotong University, Beijing 100044, China.}
\address[]{+ means authors contributed equally to this work.}

\begin{document}
\maketitle
\begin{abstract}
The rapid development of the low-altitude economy has imposed unprecedented demands on wireless infrastructure to accommodate large-scale drone deployments and facilitate intelligent services in dynamic airspace environments. However, unlocking its full potential in practical applications presents significant challenges. Traditional aerial systems predominantly focus on air-ground communication services, often neglecting the integration of sensing, computation, control, and energy-delivering functions, which hinders the ability to meet diverse mission-critical demands. Besides, the absence of systematic low-altitude airspace planning and management exacerbates issues regarding dynamic interference in three-dimensional space, coverage instability, and scalability. To overcome these challenges, a comprehensive framework, termed low-altitude wireless network (LAWN), has emerged to seamlessly integrate communication, sensing, computation, control, and air traffic management into a unified design.  This article provides a comprehensive overview of LAWN systems, introducing LAWN system fundamentals and performance evaluation metrics. Subsequently, we delve into the evolution of functional designs and review critical concerns surrounding privacy and security in the open-air network environment. We survey advanced artificial intelligence techniques that enhance LAWN functionality and enable increasingly autonomous operations. Finally, we present the cutting-edge developments in airspace structuring, air traffic management, and path planning, providing insights to facilitate the practical deployment of LAWNs.
\end{abstract}

\section{Introduction}

\subsection{Background}

\begin{figure*}[ht]
    \centering
    \includegraphics[width=0.9\linewidth]{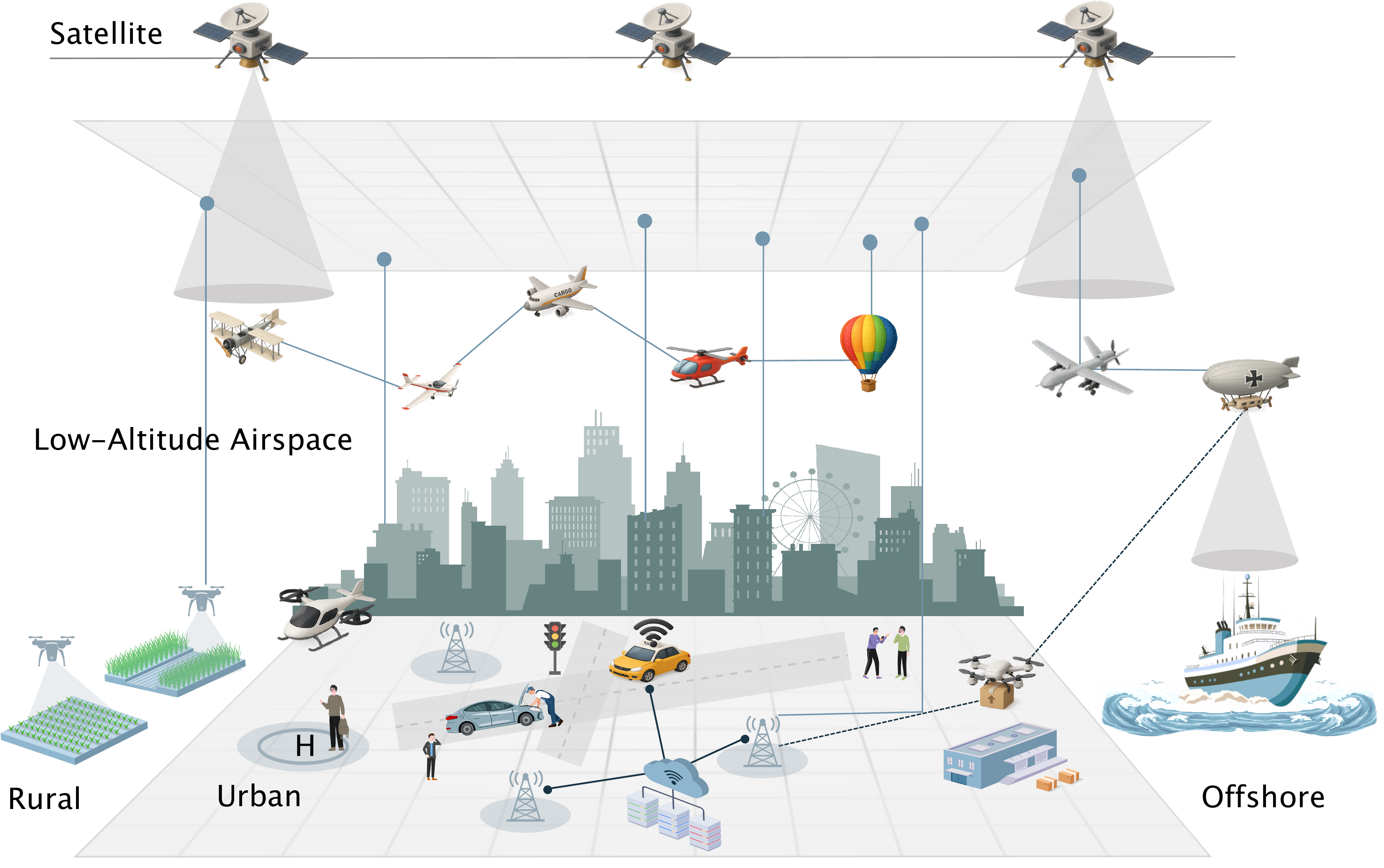}
    \caption{LAWN-enabled scenarios across rural, urban, and offshore environments with satellites and low-altitude platforms collaboratively providing ubiquitous communication, sensing, and service coverage to diverse applications.  }
    \label{figsce}
\end{figure*}
\textcolor{black}{Traditional terrestrial networks provide foundational capabilities like communication, localized sensing (e.g., global navigation satellite system (GNSS) and cellular triangulation), and fixed-point computation, supporting ubiquitous connectivity and basic data services \cite{asadi2014survey}. However, these existing networks are proving increasingly inadequate to meet the unique demands of emerging low-altitude applications. While terrestrial networks excel at their original design goals, their inherent limitations, such as static infrastructure and limited sensing capabilities, restrict their ability to effectively support dynamic and mobile sensing, communication, computation, and control in three-dimensional airspace.
This inability to adequately support low-altitude applications stems from several key limitations. First, terrestrial networks primarily offer static sensing and communication, failing to provide the dynamic, high-resolution, and mobile sensing vital for many low-altitude use cases. This makes them unsuitable for applications that require real-time environmental monitoring in diverse locations, on-demand data collection in hard-to-reach areas, or physical manipulation at varying altitudes \cite{wang2019convergence}. Second, the performance of low-altitude applications often depends on balancing complex, multi-objective requirements that cannot be managed using finite-level performance metrics. Terrestrial networks face major challenges to find end-to-end optimization that simultaneously balances data timeliness, terminal energy, and network scalability \cite{wang2025bridging}, especially in the dynamic and often unpredictable conditions of low-altitude environments. Moreover, the open nature of wireless channels makes terrestrial networks susceptible to malicious attacks. Relying on single-dimensional security and privacy measures, they are highly vulnerable to sophisticated, multi-dimensional attacks that compromise data integrity, breach privacy, cause denial-of-service, or prevent trusted communication \cite{mahmood2019analysis}. These vulnerabilities become particularly acute in low-altitude applications, where security is paramount for safety-critical operations.}

\textcolor{black}{With the rapid emergence of a low-altitude digital services ecosystem, low-altitude wireless networks (LAWNs) are increasingly recognized as essential infrastructure for building three-dimensional intelligent airspace. As illustrated in Fig.~\ref{figsce}, LAWNs are designed to provide integrated sensing, communication, computation and control services for aircraft (drones, electric vertical take-off and landing (eVTOLs), etc.) operating in the airspace below 3000 meters above ground level \cite{wang2025toward}. The core mission of LAWNs is to restructure the airspace operation paradigm through wireless technology, thereby supporting the safe and efficient operation of the low-altitude economy (LAE) \cite{liao2024benefits}, alleviating the strain on terrestrial network development, and offering technical insights for their further evolution. The detailed insights are listed as follows: }

\begin{itemize}
\item \textcolor{black}{LAWNs achieve a synergistic integration of sensing, computing, and control within a network of mobile unmanned aerial vehicles (UAVs) \cite{yang2024embodied}. This distributed approach allows for on-demand, high-resolution, and mobile environmental perception, going beyond the capabilities of static terrestrial sensors. Furthermore, it enables payload delivery or infrastructure inspection, supporting a wider range of low-altitude use cases. This distributed and dynamic sensing and interaction surpasses the limitations of static and restricted sensing/interaction characteristics of terrestrial networks, providing crucial support for low-altitude applications.}
\item \textcolor{black}{LAWNs excel in delivering diverse services to UAVs, offering far more flexible configurations than terrestrial networks can provide \cite{wang2025toward}. As illustrated in Fig.~\ref{figsce}, they dynamically adapt parameters such as UAV altitude, speed, trajectory, communication settings, and computational offloading strategies. This enables the simultaneous enhancement of data timeliness, energy efficiency, network coverage, communication reliability, and operational safety. This adaptable, system-level performance is often unattainable with the statically optimized configurations of terrestrial infrastructure, and unlocks new capabilities for aerial vehicles.}
\item \textcolor{black}{Security is a paramount concern for LAWNs, as safe and reliable operation is a prerequisite to enable a robust low-altitude economy. To this end, LAWNs leverage advanced attack defense and privacy preservation strategies, capitalizing on their mobile and distributed architecture to enable dynamic deployment of security mechanisms \cite{cai2025secure}. These include adaptive jamming resistance, secure multi-drone collaboration for threat detection, cutting-edge encryption protocols, and privacy-preserving data processing techniques. This adaptable approach yields robust, multi-dimensional security and privacy solutions specifically tailored to the unique vulnerabilities and operational contexts of aerial platforms, ensuring the safe and reliable operation of low-altitude services.}
\end{itemize}

\begin{table*}[t]
\centering
\caption{Summary of Core Contributions and Technical Focus}
\footnotesize
\begin{tabularx}{\linewidth}{|c|X|X|X|}
\hline
 \textbf{Ref.} & \textbf{Research Area} & \textbf{Core Contribution} & \textbf{Technical Focus} \\ \hline
\rowcolor{blue!10} 
\cite{jin2025advancing} & Robust Control & Overview of wireless control issues in LAWNs & MPC, remote estimation, and reliable communications \\ \hline
\rowcolor{blue!10} \cite{Work2009} & Air traffic flow management & Propose convex air traffic flow optimization models & Air traffic flow, optimization, modeling, traffic management \\ \hline
\rowcolor{blue!10} \cite{juntong2025low} & Airspace management & Investigate the LAA management &Airspace management \\ \hline
\rowcolor{green!10} \cite{yang2024embodied, kurunathan2023machine} & AI-enabled LAWN & Investigate learning-based techniques for efficient data processing & Embodied AI, machine learning, and reinforcement learning \\ \hline
\rowcolor{green!10} \cite{wang2025toward} & LAWN architectures & Discuss the realization of LAWNs covering architecture, technologies & Airspace management and energy optimization \\ \hline
\rowcolor{green!10} \cite{cordill2025comprehensive, al2025privacy, khan2024security} & Privacy and security preservation & Review privacy and security challenges in LAWNs & Physical-layer privacy and security preservation \\ \hline
\rowcolor{green!10} \cite{pandey2024uav} & Energy-aware design & Provide a comprehensive survey on RF energy harvesting & RF energy harvesting and wireless power transfer \\ \hline
\rowcolor{green!10} \cite{pogaku2022uav} & RIS-aided LAWN communications & Investigate the interplay between RIS and LAWNs & RIS \\ \hline
\rowcolor{green!10} \cite{meng2023uav} & ISAC-enabled LAWNs & Summarize LAWN-enabled ISAC for 6G networks & ISAC and sensing-aided communications \\ \hline
\rowcolor{green!10} \cite{9520337} & Computation offloading & Propose a framework for LAWN services boosting edge intelligence & Edge intelligence and computation offloading \\ \hline
\rowcolor{green!10} \cite{127,119} & Channel modeling & Introduce communication architectures for LAWNs & Channel modeling, routing, and seamless handover \\ \hline
\rowcolor{blue!10} \cite{eisenbeiss2004mini} & Remote sensing mapping & Present the design and implementation of LAWNs tailored for photogrammetric image acquisition & Photogrammetry, trajectory control, and 3-dimensional reconstruction \\ \hline
\rowcolor{blue!10} \cite{liew2017recent} & Aerial robotics & Analyze the evolution and state-of-the-art of aerial robot systems & Navigation and control \\ \hline
\rowcolor{blue!10} \cite{sdoukou2025survey} & Hardware design & Provide a comprehensive survey of UAV hardware architecture & Sensor integration and system stability \\ \hline
\rowcolor{blue!10} \cite{bouguettaya2022deep} & Image processing & Survey the image processing applications in intelligent agriculture & Deep learning \\ \hline
\rowcolor{gray!20} \textbf{Our paper} & \textbf{Both in functional designs and practical implementations} & \textbf{A comprehensive review of LAWNs in sensing, communication, computation, control, and EH; Integration of airspace management and air traffic flow optimization} & \textbf{ISAC, edge intelligence, wireless control, wireless power transfer, path planning, and air traffic management} \\ \hline
\end{tabularx}
\label{tab}
\end{table*}

\begin{figure*}[t]
  \centering
  \includegraphics[width=0.8\linewidth]{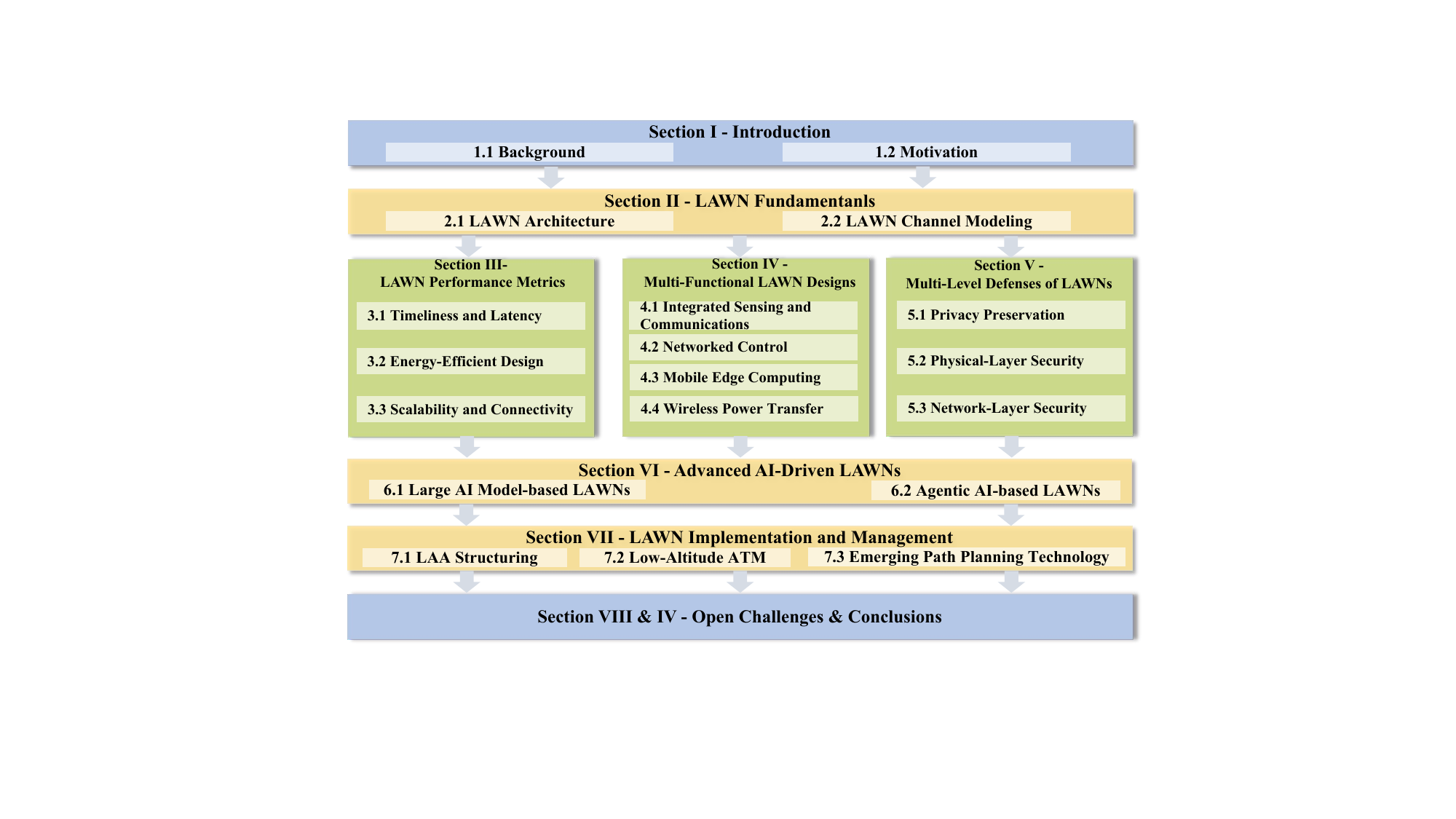}\\
  \caption{{The structure of the survey.}} \vspace{-0.3cm}
  \label{APPP}
\end{figure*}

\begin{table}[ht]
\caption{{List of Acronyms}}
\centering
\footnotesize
\setlength{\tabcolsep}{6pt}
\renewcommand{\arraystretch}{1.1}
{
\begin{tabular}{ll}
\toprule
\textbf{Acronym} & \textbf{Description} \\
\midrule
5G         & Fifth-generation\\
6G         & Sixth-generation\\
A2A        & Air-to-air \\
A2G        & Air-to-ground \\
ADS-B      & Automatic dependent surveillance–broadcast \\
AI         & Artificial intelligence \\
ATM        & Air traffic management \\
ATC        & Air traffic control\\ 
CNPC       & Control and non-payload communication \\
D2D        & Device-to-device \\
EH         & Energy harvesting \\
eVTOL      & Electric vertical take-off and landing \\
GBS        & Ground base station \\
GNSS       & Global navigation satellite system \\
HAP        & High-altitude platform \\
ISAC       & Integrated sensing and communication \\
IoT        & Internet of Things\\
LAE        & Low-altitude economy \\
LAA        & Low-altitude airspace\\
LAWN       & Low-altitude wireless network \\
LiDAR      & Light detection and ranging \\
LLM       & Large language model \\
LoS        & Line-of-sight \\
LTE        & Long-term evolution \\
MARL       & Multi-agent reinforcement learning\\
MIMO       & Multiple-input multiple-output \\
mmWave     & Millimeter wave \\
MEC        & Mobile edge computing\\
MPC        & Model predictive control \\
MFD        & Macroscopic fundamental diagram \\
NLoS       & Non-line-of-sight \\
NOMA       & Non-orthogonal multiple access\\
PID        & Proportional–integral–derivative \\
RF         & Radio frequency \\
RIS        & Reconfigurable intelligent surface\\
UAV        & Unmanned aerial vehicle \\
UAM        & Urban air mobility\\
WPT        & Wireless power transfer \\
\bottomrule
\end{tabular}}
\label{tab:acronyms}
\end{table}
\subsection{Motivation}
There have been several surveys exploring the LAWN designs from various perspectives in terms of resource allocation, energy efficiency, and learning-based techniques. 
Taking an example, the studies in \cite{wang2025toward,dai2022unmanned, 119} highlighted the flexibility and on-demand connectivity offered by LAWNs, while emphasizing the advancing capabilities of security, robust sensing, communication, and computing. The authors in \cite{9520337} introduced LAWN to enhance edge intelligence by supporting edge computing and caching services simultaneously, offering solutions to service provisioning challenges in 6G networks. Regarding techniques for learning-based LAWN designs,
the works of \cite{bai2023toward,kurunathan2023machine} comprehensively examined the application of artificial intelligence (AI) to autonomous multi-UAV wireless networks (MUWN). Furthermore,  \cite{pandey2024uav} focused on energy harvesting (EH) techniques for LAWNs, elaborating how EH can extend LAWN sustainability and facilitate connectivity for energy-constrained devices, particularly in remote or disaster-stricken areas. 
Recent surveys resorted to exploring the interplay between LAWN and key technologies for 6G. For instance, the study in \cite{pogaku2022uav} detailed the integration of reconfigurable intelligent surfaces (RIS) to enhance LAWN system performance. With the ability to manipulate signal transmission, RIS-based LAWN can improve UAV limitations in areas like coverage extension, signal reliability, and energy efficiency. The authors in \cite{meng2023uav} investigated the integrated sensing and communication (ISAC) design LAWN, demonstrating the superiority in inference management and target tracking. Beyond communication-centric research, efforts have also been devoted to multiple complementary dimensions of LAWNs. For instance, the study in \cite{jin2025advancing} evaluated various control methods of LAWNs in dynamic and uncertain environments, including model predictive control (MPC) and proportional-integrative-derivative (PID). Besides, the study in \cite{sdoukou2025survey} discussed the fundamental hardware components of multi-copter UAVs, outlining key methods for optimizing critical parameters to ensure reliable and effective UAV usage.  The work in \cite{Work2009} further summarized convex formulations for air traffic flow optimization, presenting a solution to addressing challenges such as congestion and capacity constraints.

\textcolor{black}{For ease of presentation}, we compare the related surveys in \textbf{TABLE} \ref{tab}. It can be seen that on the one hand, the rise of LAWNs can bring plenty of advantages such as improved coverage in underserved areas, flexible deployment options, and enhanced line-of-sight communications. On the other hand, due to the challenges of high-volume, multi-connection data, and the open nature of wireless channels, key concerns in terms of reliable communications, accurate sensing, and robust control need further joint consideration.
It is, however, noted that the related surveys do not fully consider the multiple functions, various indicators, and intertwined privacy and security issues of the LAWNs holistically. Moreover, the practical implementation of LAWNs requires an investigation about the low-altitude airspace (LAA) structuring and management. In this regard, existing literature with limited scopes of LAWNs may not provide a complete understanding of the trade-offs between performance, functionality, and security/privacy, and potentially result in suboptimal designs and deployments for real-world LAWN applications. To close up the above research gap, this article provides an overview of LAWN from functionality and performance evaluation, as well as air traffic management (ATM), offering a more comprehensive and broader perspective on the design, deployment, and management of  LAWN systems.
The main contributions of the paper can be summarized as follows:

\begin{itemize}
\item \textcolor{black}{We demonstrate the necessity of LAWNs for enabling the growth and sustainability of the low-altitude economy, showcasing their superiority over traditional terrestrial networks in offering enhanced functionality, satisfying more stringent performance demands, and providing more secure operating environments. Specifically, LAWNs can deliver on-demand connectivity, dynamic resource allocation, and resilient communication links in scenarios where terrestrial infrastructure is limited or unavailable, thereby unlocking new possibilities for aerial services and economic activities, while simultaneously offering advanced security features to mitigate vulnerabilities inherent in wireless communications.}
\item We define the concept of LAWNs and present the  foundations from architectural, functional, and technical perspectives. On this basis, key questions about the nature of LAWNs, their architectural design, and their inherent operating mechanisms can be addressed. Furthermore, by outlining these fundamental aspects, we provide a comprehensive framework for understanding the potential and limitations of LAWNs in various applications.
\item We offer insights into LAWNs from functional, performance, privacy, and security viewpoints. Specifically, focusing on sensing, computing, communication, and control aspects, we illustrate how LAWNs enable diverse functionalities, optimize critical performance indicators, effectively protect user privacy, and robustly defend against malicious attacks. This holistic analysis provides a comprehensive understanding of the benefits and challenges associated with deploying LAWNs in real-world environments.
\end{itemize}

The remainder of the paper is structured as shown in Fig.~\ref{APPP}. Section~\ref{sec2} introduces the fundamentals of LAWNs, including system architecture and representative channel models. Then, section~\ref{sec3} discusses LAWN performance metrics in terms of timeliness and latency, energy-efficient design, scalability, and connectivity. In section~\ref{sec4}, we present multi-functional LAWN design methods. Subsequently, section~\ref{sec5} is devoted to multi-level defenses of LAWNs, covering privacy preservation, physical-layer security, and network-layer security. Section~\ref{sec6} focuses on advanced AI-driven LAWNs, including both large-model-based and agentic AI-based architectures. Section~\ref{sec7} addresses implementation and management aspects in terms of airspace structuring, traffic management, and path planning. Finally, the open research challenges and conclusions are outlined in sections~\ref{sec8} and \ref{sec9}, respectively. For brevity, we further summarize the acronyms of this survey in \textbf{TABLE \ref{tab:acronyms}}.

\section{LAWN Fundamentals} \label{sec2}
\subsection{LAWN Architecture}

\begin{figure*}[t]
    \centering
    \includegraphics[width=\linewidth]{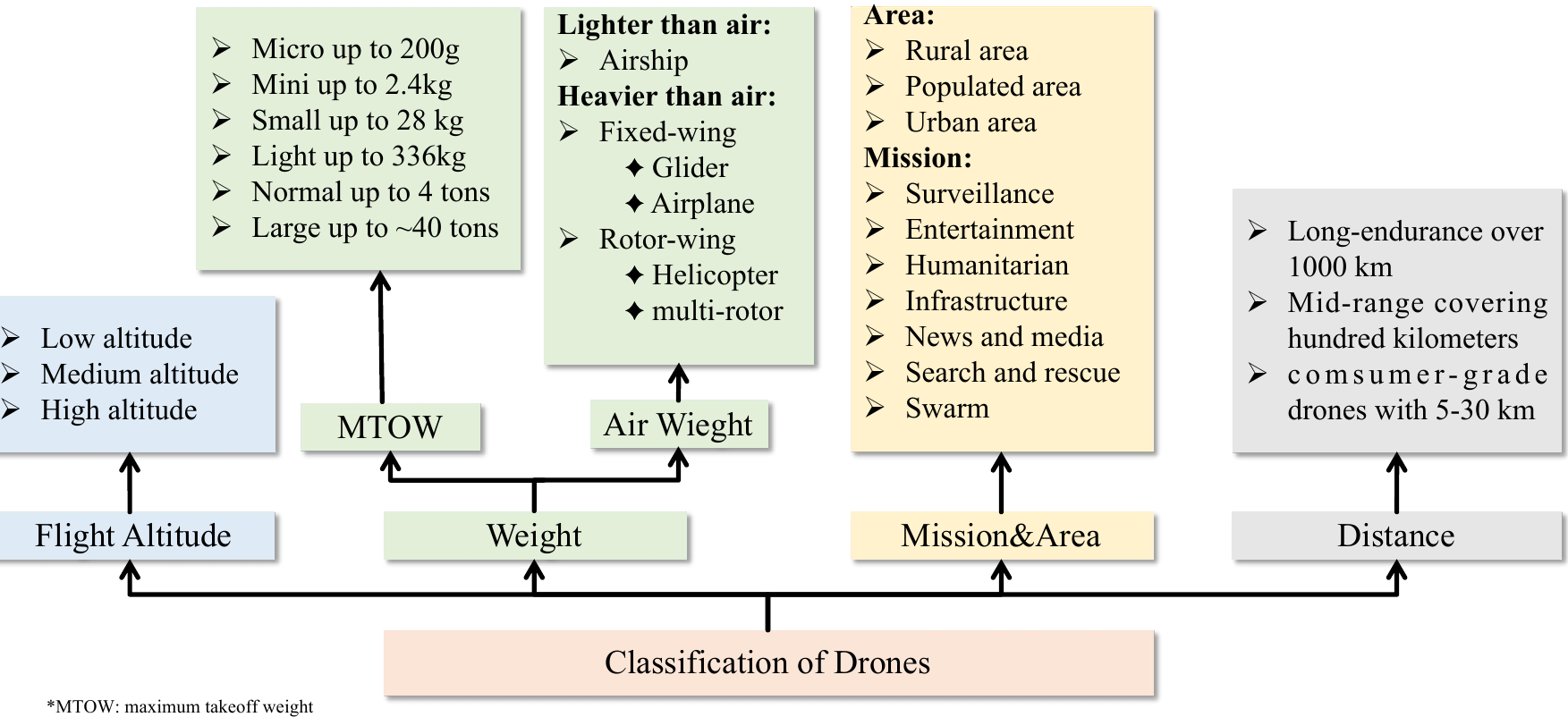}
    \caption{Classification of drones under various criteria.}
\label{fig:drone_classification}
\end{figure*}

In this subsection, we first introduce the concept of LAWN, emphasizing its core components and key distinctions from conventional UAV-enabled communication networks, followed by presenting the architectures. LAWN represents a next-generation infrastructure paradigm designed to support the emerging low-altitude economy, encompassing a diverse set of aerial platforms operating within airspace below 3000 meters. Rather than simply extending terrestrial cellular infrastructure into the aerial domain, LAWN goes beyond communication connectivity to support dense, diverse, and task-driven aerial nodes with integrated functions of communication, sensing, control, and service delivery.
Specifically, a canonical LAWN consists of the following tightly coupled components:

\textit{1) Aerial platform}: The aerial platforms form the physical backbone of LAWNs, providing the carriers for wireless control, distributed sensing, and service delivery. Following the classification in \cite{atkins2009commercial,dalamagkidis2008unmanned}, these platforms can be distinguished along several complementary dimensions. In terms of flight altitude, operations typically span low-, medium-, and high-altitude regimes within the sub-3000~m airspace. From a weight perspective, platforms range from micro to large aircraft and include both lighter-than-air vehicles and heavier-than-air fixed/rotor-wing designs. Mission profile and deployment area further differentiate platforms used in rural, suburban, and dense urban environments for tasks such as surveillance, entertainment, humanitarian relief, and swarm operations. Finally, range and endurance give rise to long-endurance systems capable of flights beyond 1000~km, mid-range platforms covering several hundred kilometers, and consumer-grade drones that typically operate within about 5–30~km under remote or app-based control. To sum up, the classification of drones under various criteria is demonstrated in Fig.~\ref{fig:drone_classification}.

 \textit{2) Terrestrial Infrastructure}: The terrestrial infrastructure forms the backbone to support practical LAWN realizations. In particular, the takeoff and landing places like vertiports are equipped with necessary turnaround supports for battery charging/refueling and repairing. Meanwhile, the ground base stations (GBS) are responsible for data transmission conveying mission planning. In addition, the environmental services, including geofencing, high-definition maps, and weather forecasts, provide crucial inputs for predictive designs. 

\textit{3) Low-Altitude Airspace Management System}: The airspace management system is critical for ensuring efficient operations within LAWNs by coordinating flight paths, altitude assignments, and timing to prevent conflicts between drones, thereby maintaining safety even in congested airspace. Relying on automatic dependent surveillance-broadcast (ADS-B) and GNSS-based localization to monitor air traffic, advanced de-confliction algorithms dynamically adjust flight routes to account for changing environmental conditions and unexpected events. Moreover, by integrating with traditional aviation control systems, the LAA management framework facilitates seamless coordination between unmanned and manned aircraft, ensuring smooth integration within existing civil aviation infrastructure.

{
It is evident that aerial-to-ground (A2G) communications are indispensable functions within LAWNs. However, we herein highlight that while LAWN systems are established on UAV-based platforms, they differ fundamentally from conventional UAV communication architectures. Therefore, it is critical to clarify the scope of LAWNs to avoid confusion. Functionally, traditional UAV communication systems emphasize wireless transmission performance at the physical layer, where aerial nodes typically act as terminal users or relays within pre-established cellular or Ad-hoc networks, with minimal focus on flight attitude control, battery management, and formation coordination. In contrast, LAWNs extend beyond traditional communication purposes by supporting the seamless integration of sensing, computation, and control, enabling real-time coordination across multiple domains. It is not intended to replace dedicated aerial communication services, which may still be offloaded to conventional networks. Instead, LAWN focuses on the co-design of networking models, control-aware protocols, spatial scheduling, and real-time decision-making across airspace-aware nodes. Moreover, it emphasizes large-scale network coordination, enabling collaborative efforts across multiple airspace-aware nodes to optimize performance and efficiency \cite{yuan2025ground}. Furthermore,  LAWNs require comprehensive optimization beyond physical layers. For instance, efficient routing and network protocols must be co-designed to accommodate dynamic drone mobility and task-driven communication needs, while application layer designs ensure that mission-critical tasks can be real-time controlled by remote apps.}

{
From a broader networking perspective, LAWNs are also conceptually distinct from space-air-ground integrated networks (SAGINs), which are designed to provide global coverage by vertically integrating satellite, aerial, and terrestrial segments into a unified communication infrastructure \cite{chen2024survey}. SAGIN research primarily concentrates on multi-layer connectivity, backhaul routing, and resource sharing across heterogeneous layers, while low-altitude airspace management, mission-centric control, and interaction with urban environments are only indirectly addressed. LAWNs, on the other hand, focus on the sub-3000 m domain and are tailored for dense, task-driven operations that tightly couple airspace structuring, traffic management, sensing, computation, and control. Rather than being regarded as a subset of SAGIN, LAWNs are better interpreted as a complementary low-altitude operational layer that can interconnect with SAGIN for backbone connectivity, while maintaining control-aware protocols and airspace-aware service for LAE applications.
}

\begin{table*}[t]
{  \centering
    \footnotesize
    \caption{{Comparison of representative LAWN architectures}}
    \label{tab:lawn_topology_comparison}
    \setlength{\tabcolsep}{4pt}
    \renewcommand{\arraystretch}{1.1}
    \begin{tabular}{m{2.2cm} m{3cm} m{3.2cm} m{3.5cm} m{4cm}}
  \toprule
        \textbf{Topology} & \textbf{Reliability} & \textbf{Latency} & \textbf{Resource efficiency} & \textbf{Scalability} \\
        \midrule
        Star network &
        Vulnerable to central-node failure&        Low for nearby nodes, grows with hub load and distance &
        High for small-scale networks with simple centralized routing &
        Limited; performance degrades as the node number increases \\
        \midrule
        Mesh network &
        Moderate; enabling local rerouting under failures &
        Moderate due to multi-hop forwarding &
        Moderate; incurring additional signaling and energy cost &
        Good for medium-size deployments \\
        \midrule
        Multi-group mesh &
        High; backbone links connecting groups &
        Low for intra-group; moderate for inter-group  &
        Improved efficiency through balanced load across groups &
        High; partitioning into groups supports large systems with manageable overhead \\
        \midrule
        Hierarchical mesh &
        High; tiered redundancy and fault isolation &
        Low; high-tier links offer short end-to-end paths for wide-area communication &
        High; aggregated control and traffic management enhance global resource usage &
        Very high; suitable for large and heterogeneous LAWNs with diverse service demands \\
        \bottomrule
    \end{tabular}    }
\end{table*}

\begin{figure}
    \centering
    \includegraphics[width=\linewidth]{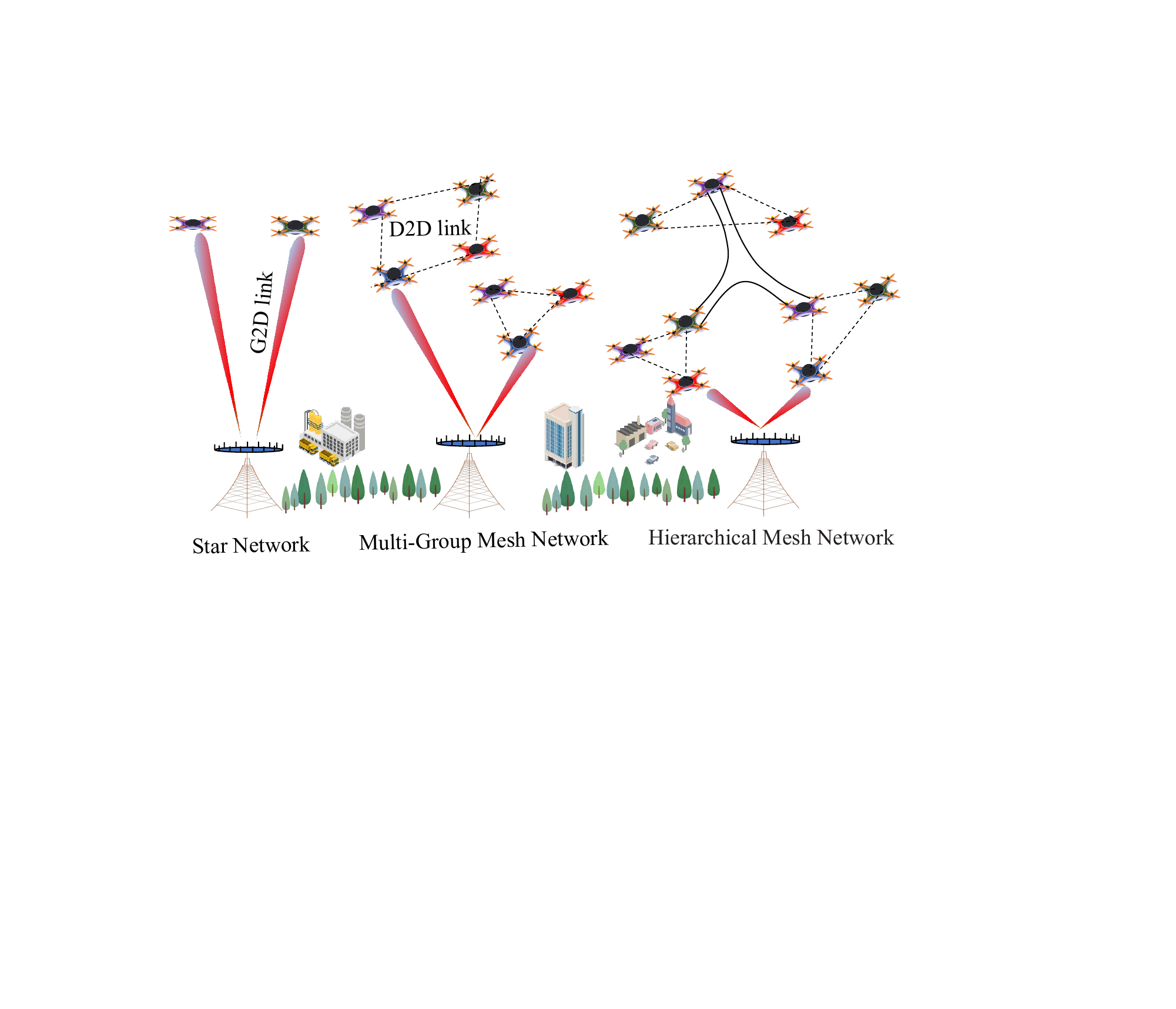}
    \caption{LAWN architectures, including star networks with direct A2G links, multi-group mesh networks with intra-group cooperation, and hierarchical mesh networks with multi-layer A2A connections for scalable coverage.}
    \label{figarchr}
\end{figure}

To ensure the effective functioning of LAWNs, it is essential to establish seamless and efficient information flow, both between the ground crew and UAVs and amongst the UAVs themselves. This, in turn, demands a robust and adaptive network architecture that can support not only high-capacity data transfer but also real-time coordination of various operational elements. Consequently, we proceed to focus on LAWN architectures ranging from star to mesh networks shown in Fig. \ref{figarchr}, particularly emphasizing multi-group and hierarchical mesh networks \cite{119,akyildiz2005wireless}.

\textit{1) Star Network:} In a star network, all LAWN nodes are connected to a single central ground station via dedicated A2G links, making it a straightforward and efficient solution for small-scale operations. The central station controls and coordinates the communication flow between the nodes. While this centralized architecture is easy to deploy, it presents significant limitations as the network grows. The primary issue is the single point of failure associated with the central ground station. If the ground station fails, the entire communication network is disrupted. Moreover, as LAWN nodes move farther from the central node, the communication delay and bandwidth limitations become more problematic, rendering star networks inefficient for large, dynamic missions.

\textit{2) Mesh Networks: }
As the scope and complexity of LAWNs continue to expand, the inherent limitations of star network topologies become increasingly evident. To overcome these limitations, mesh networks provide a decentralized communication architecture. In a mesh network, each LAWN node can communicate directly with other nodes via air-to-air (A2A) links, either via direct communication or through multiple hops. The decentralized structure ensures that even if one node fails or moves out of range, the network can adapt and maintain connectivity. As a consequence, mesh networks are particularly well-suited for LAWN systems to maintain robustness and reliability in high-mobility environments.
Ad-hoc networks naturally form as a subset of mesh networks, where each LAWN node autonomously communicates with nearby nodes, establishing a peer-to-peer communication model. As discussed in \cite{120}, the flexibility of Ad-hoc networks ensures that LAWN nodes can adapt to changes in topology and network conditions, enabling continuous communication even in remote areas where communication infrastructure is unavailable. While a single mesh network offers significant improvements over star networks, the multi-group mesh network architecture provides an effective solution for larger LAWN systems. In a multi-group mesh network, multiple independent mesh networks operate in parallel, each serving a subset of LAWN nodes. Communication between these groups is facilitated by backbone nodes that connect the different mesh networks, ensuring that data can flow seamlessly across the entire LAWNs. Moreover, compared to a single mesh network, multi-group networks allow for efficient load balancing, reducing congestion and improving overall network performance.
For even larger systems, hierarchical mesh networks offer an efficient solution \cite{127}. In a hierarchical mesh network, nodes are organized into several hierarchical layers, each responsible for different communication tasks. The lower layers consist of local mesh networks, where LAWN nodes communicate within their group. Higher layers, composed of more powerful nodes, manage inter-group communication and connect the network to the ground station or other higher-tier nodes. This hierarchical structure enhances the network’s scalability and fault tolerance by distributing the communication load across different layers.
The multi-layer design also improves network efficiency by optimizing communication paths, reducing congestion at the higher layers, and ensuring that the network remains operational even if one or more layers fail. The hierarchical approach is particularly beneficial for missions involving large numbers of LAWN nodes with varying communication capabilities, such as self-organized drone formations working in coordination for search and rescue operations or wide-area surveillance. To sum up, the representative LAWN architectures discussed above are compared in \textbf{TABLE}~\ref{tab:lawn_topology_comparison} in terms of reliability, latency, resource efficiency, and scalability.

\subsection{LAWN Channel Modeling}

Channel modeling serves as the cornerstone for ensuring safe and reliable UAV operations, as it captures the propagation behaviors that underpin performance evaluation and system design. In this context, spectrum allocations define the operating conditions that directly shape the propagation characteristics. To this end, the International Civil Aviation Organization (ICAO) requires that control and non-payload communication (CNPC) links operate in protected spectrum bands. In line with this requirement, the International Telecommunication Union (ITU) has designated specific portions of the L-band (960–977 MHz) and C-band (5030–5091 MHz) for CNPC services in UAV systems \cite{khuwaja2018survey}. For beyond-line-of-sight (LoS) operations involving satellite communication, the Ku-band and Ka-band have been reserved under the aeronautical safety spectrum framework to support robust satellite-based CNPC connectivity. In addition, national regulators, such as those in China, have authorized supplementary LoS frequency bands to meet the rising demand for UAV deployments in densely utilized low-altitude airspaces \cite{luo}.
Based on the allocated bands, traditional studies on aeronautical channel modeling primarily target high-altitude, long-range aircraft operating under transcontinental or military flight conditions. The work in~\cite{bello1973aeronautical} introduced one of the earliest statistical frameworks for such environments, analyzing Doppler spread and multipath dispersion caused by rapid airborne motion. In~\cite{haas2002aeronautical}, the authors categorized modeling approaches into deterministic, stochastic, and geometry-based types, highlighting the influence of aircraft altitude, elevation angle, and antenna orientation on channel characteristics such as path loss and Ricean K-factors. The authors in~\cite{jain2012requirements} further examined wireless datalink requirements for large-scale unmanned systems, analyzing the trade-offs among various candidate technologies under high-mobility and limited-spectrum conditions.  While primarily focused on conventional high-altitude aircraft, the resulting models and statistical parameters offer a foundational reference for extending channel characterization into LAWNs.

The A2G channel is pivotal for facilitating data exchange within LAWNs, making accurate modeling of the A2G propagation channel essential for the design of resilient drone-based communication systems. In urban environments where LAWNs are commonly deployed, the propagation channels are often shaped by dense building structures, irregular street canyons, and pronounced multi-path effects, leading to significant challenges for traditional propagation models. A representative statistical modeling approach is presented in \cite{6492100}, where the authors conducted extensive UAV field measurements at low elevation angles across urban environments. They developed a time-series generator capable of capturing the elevation-dependent fading characteristics, providing a realistic stochastic representation of mobile A2G links for simulation and system design. 
Furthermore, the authors in \cite{xu20223} studied 6G low-altitude mmWave massive multiple-input multiple-output (MIMO) A2G channels in urban scenarios, capturing space-time-frequency non-stationarity with UAV-specific parameters.
To further quantify large-scale shadowing behavior of A2G channels, an elevation-dependent shadowing model was proposed in \cite{4483593}. Originally developed for high-altitude platforms (HAPs), the model remains applicable to LAWNs operating over urban areas. The key insight is that the shadowing loss exhibits a strong dependence on the elevation angle, following an exponential trend as the UAV moves closer to the horizon, thus affecting LoS probability and average signal strength. The work of \cite{7037248} investigated an analytical A2G path loss model that integrates stochastic geometry with ray-tracing results, which separates LoS and non-LoS (NLoS) components and incorporates a probabilistic LoS function based on elevation angle and building distribution.
{
Beyond classical stochastic and geometry-based models, data-driven channel prediction offers a complementary way to address the strong nonstationarity of low-altitude links. In the hybrid framework of \cite{8663422}, a physics-based state evolution model captures large-scale effects such as elevation-dependent path loss and shadowing, while a learned predictor trained on measurement data models small-scale fading, blockage events, and environment-specific dynamics that are difficult to describe analytically. The combined use of model-based state evolution and data-driven learning is particularly suitable for LAWNs, enabling accurate short-term channel forecasting for UAV relays operating in complex urban environments.}

In addition to urban deployments, the study in \cite{8048502} proposed an analytical path-loss model for suburban A2G transmissions. By incorporating UAV height and LoS probability, it facilitates efficient planning of UAV trajectories and resource allocation in semi-urban environments. Regarding over-water scenarios, the authors in \cite{5089553} developed a geometry-based multi-path model capturing sea surface reflections and long-range propagation delays for A2G links. Meanwhile, the work in \cite{5741881} conducted C-band measurements over sea surfaces at low altitudes, characterizing channel fading patterns and multi-path effects essential for UAV telemetry and maritime missions. Notably, LAWN systems are increasingly integrated into cellular networks, which exhibit distinct propagation characteristics. The work in \cite{7936620} conducted empirical studies of UAV communication over commercial long-term evolution (LTE) networks, which revealed that drones flying above the downtilted beam of base station antennas are likely to result in higher path loss and interference levels. The authors in \cite{8606910} further explored mid-band 5G service delivery for ultra-low-altitude UAV base stations, whose measurements demonstrated significant Doppler effects and elevation-dependent delay profiles, emphasizing the importance of precise modeling for 5G aerial deployments. To complement theoretical modeling, a series of wideband channel measurement campaigns were conducted in~\cite{matolak2017air1, sun2017air2, matolak2017air3}, targeting A2G links over diverse terrains including over-water, mountainous, and suburban areas, providing empirical models capturing delay spread, angular dispersion, and Ricean fading statistics with varying antenna heights and aircraft velocities. 
\begin{table*}[t]
{\caption{{Summary of representative LAWN channel modeling }}
\centering
\footnotesize
\setlength{\tabcolsep}{6pt}
\renewcommand{\arraystretch}{1.12}
\begin{tabular}{p{3.6cm}p{3.5cm}p{3cm}p{3.4cm}p{1.6cm}}
\toprule
\textbf{Scenario} & \textbf{Frequency band} & \textbf{Link} & \textbf{Aircraft type}  & \textbf{Ref.} \\
\midrule
Urban (100–170 m altitude) & 2 GHz & A2G & Airship & \cite{6492100} \\
Dense urban & MmWave & A2G/A2A  & Low-altitude UAVs & \cite{xu20223} \\
Urban & Sub-6 GHz & A2G & HAP & \cite{4483593,7037248} \\
Over-water / maritime & C band & A2G & Small UAVs & \cite{5089553,5741881} \\
Rural/mountain/near-urban & L/C bands & A2G & UAVs & \cite{matolak2017air1,sun2017air2,matolak2017air3} \\
Suburban / rural & LTE band & A2G & UAVs & \cite{7936620} \\
Open field & 2.4/5 GHz (IEEE 802.11) & A2G & Fixed-wing UAVs & \cite{4086461} \\
Terrestrial and maritime & C band & A2A & Small fixed-wing UAVs & \cite{7991486} \\
Aerial relay & Sub-6 GHz & A2A & UAVs & \cite{dac.1212} \\
Open field/sea & VHF/L/C bands & Air-to-space/A2A/A2G & Civil aircraft & \cite{bello1973aeronautical,haas2002aeronautical,jain2012requirements} \\
\bottomrule
\multicolumn{5}{l}{*VHF: very high frequency (30–300 MHz); L-band (960–977 MHz); C-band (5030–5091 MHz) }\\
\end{tabular}
\label{lawn_channel_unified_compact}}
\end{table*}
Apart from A2G channels, A2A communication channel modeling is equally critical within LAWNs, which emerges in scenarios involving aerial relaying, drone swarms, and autonomous formation control, exhibiting fundamentally distinct propagation behaviors compared to A2G links. Specifically, while A2G channels are often subject to shadowing and ground reflections, A2A links generally operate under more favorable LoS conditions with fewer multi-path effects. However, A2A communication is highly influenced by the symmetric mobility of both transceivers, resulting in dynamic spatial-temporal variations in the channel characteristics. The work in \cite{7991486} presented a comprehensive C-band measurement involving dual fixed-wing UAVs to characterize A2A propagation in both terrestrial and maritime environments, in which the analysis reveals that LoS paths dominate in most scenarios, with secondary reflections arising from ground or sea surfaces. Notably, the maximum observed path delay spread remains within 4\,$\mu$s at flight altitudes below 700\,m, providing practical design insights for CNPC links. Additionally, the study emphasizes that elevation angle and terrain type significantly influence the presence and power of multi-path components, particularly at low altitudes or over undulating terrain. In parallel, the analytical framework presented in \cite{dac.1212} explored A2A relaying within Nakagami-$m$ fading channels, incorporating Doppler shifts induced by mobility and the concept of cooperative diversity. The authors assessed multi-carrier cooperative relay strategies under diverse fading conditions, demonstrating how relative motion and angular alignment between UAVs influence both outage probability and spectral efficiency. The study underscores the necessity for statistical models that comprehensively account for fading severity, mobility dynamics, and relay topology. Furthermore, the work in \cite{ahmed2016importance} emphasized the pivotal role of precise link characterization in A2A communications, essential for ensuring reliable transmission in dynamic environments. It was shown that environmental factors, antenna orientation, and multi-path fading substantially affect communication link quality. The authors highlighted that protocols based on overly simplistic or inaccurate models often fail in real-world scenarios, advocating for empirical link characterization to improve the design of A2A communication protocols.
From a modeling perspective, A2A channels pose unique challenges, where the bilateral mobility of transmitter and receiver introduces dual Doppler shifts and rapidly varying link geometries, making static or ground-referenced models inapplicable. Moreover, the absence of terrain-induced shadowing or rich scattering often leads to sparse multi-path profiles with limited angular dispersion, necessitating alternative approaches such as geometry-based stochastic modeling or real-time measurement-driven adaptation. {To facilitate a comprehensive understanding of the LAWN channel models, a unified comparison is presented in \textbf{TABLE} \ref{lawn_channel_unified_compact}, which consolidates key parameters, including frequency bands, link types, and aircraft types, across different implementation scenarios.}

\section{LAWN Performance Metrics}\label{sec3}
In this section, we investigate key performance metrics for LAWN designs, focusing on latency, energy efficiency (EE), and scalability.
\subsection{Timeliness and Latency}

In practice, LAWNs are frequently deployed to support time-sensitive applications, where the transmission delay may lead to severe operational and safety implications. To name a few, in emergency response, the timely delivery of situational data such as survivor locations, fire spread, or hazardous material detection is essential for enabling rapid decision-making and coordinated rescue actions. Therefore, one of the core design principles for LAWNs is the minimization of end-to-end latency while maintaining reliable data transmission. To achieve this, finite-block length (FBL) transmission has emerged as an important technique for managing communication in constrained networks. Unlike the traditional Shannon capacity framework, which assumes infinitely long codewords, FBL transmission provides a more practical characterization of the achievable rates given block lengths and error probability. More importantly, the channel conditions in LAWNs often change rapidly, leading to the reliable infinite block length transmission assumption being invalid. The authors of \cite{zhou2023finite,zhou2017second,zhou2019non} investigated finite blocklength multiterminal lossy source coding, providing new benchmarks for efficient compression with multiple encoders, decoders, and source sequences, thereby strengthening the theoretical foundation of multiterminal low-latency communications and greatly promoting the development of low-altitude low-latency networking. The work of \cite{9126811} investigated IoT service provisioning by leveraging FBL transmission in reducing communication delays within drone-aided networks, with drones being both communication relays and computational resources. Moreover, Jin \textit{et al.} proposed the co-design of sensing, communication, and control in LAWNs, where FBL transmission is utilized with remote control tasks to minimize latency. The results show that optimizing resource allocation for communication and control can lead to lower delays and better overall performance \cite{11045436}.

Beyond FBL transmission, some recent studies focus on low-latency communications. For instance, the authors in \cite{liu2023predictive} presented a predictive precoder design for ultra-reliable low-latency communication (URLLC) systems, where a deep learning (DL)-based framework was developed to predict the precoder via historical channel data. By leveraging a convolutional long short-term memory (CLSTM) network, reliable communication with minimal latency can be guaranteed, which demonstrates superior frame error rate (FER) performance approaching the theoretical lower bound with reduced latency. More relatively, the work of \cite{cai2022resource} explored resource allocation for a two-way UAV relaying system under URLLC constraints, focusing on minimizing latency in both the forward and backward links. An iterative approach was proposed to maximize the system throughput while fulfilling the latency and decoding error probability requirements. 

Moreover, the age of information (AoI) is another critical metric facilitating delay-aware LAWN design.  In particular, the AoI measures the freshness of data in a communication network, where outdated data can lead to poor decision-making performance. Toward this end, the study in \cite{9151993} explored UAV-assisted wireless-powered IoT networks, which optimized UAV trajectories to minimize AoI and ensure that the sensor data remains fresh. Furthermore, Gao et al. examined AoI-sensitive data collection in multi-UAV-assisted wireless sensor networks \cite{10007855}, in which a clustering algorithm, devised to optimize both AoI and energy consumption in multi-UAV systems, was developed. The results show that by adjusting the UAV flight paths and clustering strategies, AoI can be minimized such that LAWN nodes have access to the freshest data with low energy usage, highlighting that AoI management is essential for real-time collaboration and successful task execution within LAWNs.
The importance of real-time scheduling in LAWN systems was studied in \cite{10978356} to manage communication effectively. By prioritizing the transmission of high-priority data and optimizing the transmission schedules, real-time scheduling can help minimize latency and AoI.

\subsection{Energy-Efficient Design}
The sustainability of LAWNs is inherently constrained by the limited onboard energy and computational resources of UAVs, making energy-efficient design essential for long-term viability in real-world applications. Early studies primarily examined energy-aware deployment strategies to extend system lifetime. For instance, \cite{9126212} proposed energy-saving deployment algorithms for UAV swarms to optimize wireless coverage while minimizing energy consumption. In a similar approach, the study of \cite{8364586} addressed the rapid deployment problem of heterogeneous UAVs to provide wireless coverage over a target area, where the objective for minimizing the maximum deployment delay to ensure fairness and minimizing the total deployment delay to enhance efficiency were addressed, respectively. However, these works largely focus on stationary optimization and do not fully exploit the high-mobility characteristics of LAWNs. In the context, the authors in \cite{8068199} investigated joint trajectory and transmit power control for UAV relay networks to minimize the outage probability.  By carefully devising UAV trajectories and power allocation, energy consumption can be reduced without sacrificing communication reliability.  Further, the study in \cite{yang2019energy} proposed a UAV-assisted backscatter communication system for energy-efficient IoT, which shows a tradeoff between collection location and outage probability on the improvement of EE. On this basis, the work of \cite{11129138} investigated the trade-off between communication/sensing performance and energy consumption in ISAC-LAWN systems, where the goal was to maximize worst-case EE by jointly designing beamforming and UAV trajectory.

Note that \cite{9126212,8364586,8068199,yang2019energy,11129138} primarily focused on the transmit power control, significantly lower than the propulsion-related power. Hence, the work of \cite{10159441} proposed a LAWN system simultaneously sensing ground users and forwarding the sensed information. The authors formulated a multi-objective optimization problem to maximize the EE by jointly optimizing user scheduling, movement-related power, and UAV trajectory.
 While existing methods have improved network EE, their optimization objectives are often limited to reducing the energy consumption of individual UAVs, overlooking the potential gains from cooperative flight dynamics in LAWNs. In real-world deployments, UAVs inevitably interact with surrounding airflow patterns, which can be strategically exploited to boost overall EE. Specifically, airflow over UAV wings generates wake vortices that create an upwash region behind the wingtips. A trailing UAV entering this region experiences reduced aerodynamic drag, thereby lowering propulsion energy consumption, as well-documented in avian flocking. Building on this principle, the studies in \cite{weimerskirch2001energy, wu2025toward} introduced an ISAC-LAWN design that jointly takes into account aerodynamic factors and sensing/communication performance, showing that a 'V'-shaped formation is likely to be aerodynamically favorable.

It is evident that the above approaches can only save energy to extend network lifetime, lacking the ability to recharge the battery to achieve a long-term power supply. To address this challenge, the authors in \cite{10139785} explored the performance improvement of UAV communication systems integrated with solar energy harvesting. The uncertainty of solar energy supply was investigated, followed by presenting an iterative optimization approach to maximize throughput. The simulations show that integrating solar energy into LAWN systems can significantly extend the operational duration, thereby enhancing the sustainability in remote areas where recharging infrastructure is scarce. Meanwhile, the work of \cite{8941314} investigated energy-efficient LAWN communications design with energy harvesting, where trajectory optimization techniques were used to minimize energy consumption while maximizing the coverage area as well as maintaining QoS.

\subsection{Scalability and Connectivity}

Benefiting from the high mobility and on-demand deployment capabilities, scalability and connectivity are essential for LAWNs to adapt to various practical requirements. The performance of large-scale LAWN networks is heavily influenced by the ability to manage communication capacity, coverage area, and resilience under dynamic conditions, such as UAV movement and interference. For instance, to reveal the fundamental performance limits of beam alignment for mobile targets such as UAVs in low-altitude networks, the authors in \cite{zhou2021resolution,zhou2022resolution} investigated mmWave multiple-antenna communications, deriving theoretical benchmarks for optimal non-adaptive query procedures in multidimensional target search and extending them to scenarios with multiple and moving targets. They further demonstrated the benefits of adaptive query procedures, thereby providing finite-blocklength benchmarks that enable reliable connectivity in low-altitude networks.  As evidenced by \cite{10557642}, wherein network localization and formation control in LAWN systems with asynchronous agents were investigated. Specifically, the authors developed a message-passing algorithm using Gaussian messages for efficient localization and clock offset estimation to maintain robust information exchange across all LAWN nodes. Subsequently, two types of control policies, namely angle-based and distance-based formation control, were employed for adaptive obstacle avoidance. By doing so, the LAWN systems are capable of quickly responding to environmental changes, and thereby, enhancing scalability and connectivity. 

To enhance the network capacity, the authors in \cite{7451189} proposed to resort to a heterogeneous LAWN where UAVs act as aerial base stations that can dynamically adjust to varying traffic loads in 5G environments. By optimizing interference coordination and spectrum sharing mechanisms, it is shown that the proposed method can effectively increase capacity, particularly in areas with high user density. As observed, \cite{7451189} underscored the potential for LAWN systems to scale up efficiently, ensuring that as network demand increases, UAVs can provide the necessary coverage and capacity. Furthermore, the authors in \cite{9126212} optimized UAV swarm deployments aimed at providing wireless coverage by adjusting their flying distances and altitudes. As LAWN systems expand, it is crucial to ensure resilience to network disruptions. To this end, the authors in \cite{8998329} examined how LAWN systems connected to cellular networks perform under conditions with high mobility and diverse user densities, where the impact of UAV mobility on handover rates and connectivity was analyzed. To mitigate handover failure, the authors proposed coordinated multi-point (CoMP) transmission, ensuring stable connectivity and resilience even in scenarios when drones move through areas with varying signal strength.  In \cite{9893396}, the authors focused on the ISAC-LAWN designs, demonstrating that sensing capability can improve resilience by enabling the network to better adapt to disruptions in communication paths.  

However, the coexistence of heterogeneous LAWN nodes sharing spectral and spatial domains creates a complex interference environment that challenges large-scale deployment and reliable interconnection. The main challenges stem from beam pattern overlaps and waveform interactions across both spatial and temporal dimensions. To tackle these issues, the work in \cite{WU2025103451} introduced a mode-switching strategy, where UAVs alternate between jamming and information transmission. By optimizing mode selection, scheduling, and trajectories, this approach ensures strong A2G links even in the presence of mobile ground users. Furthermore, \cite{9143143} proposed a deep learning–based predictive beamforming scheme that mitigated UAV jittering by forecasting the angles between UAV and user, thereby enabling robust and reliable UAV communications. Moreover, the authors in \cite{9551699} proposed a dynamic interference management framework by jointly optimizing UAV trajectories and power allocation, thereby improving overall network efficiency under dense deployments.  Beyond intra-LAWN coordination, such solutions can also be extended to satellite integration. For example, the authors in \cite{10680056} designed robust beamforming schemes for SAGIN to guarantee reliable downlink communications. 

\section{Multi-Functional LAWN Designs} \label{sec4}

\subsection{Integrated Sensing and Communications}
\begin{figure*}
    \centering
    \includegraphics[width=\linewidth]{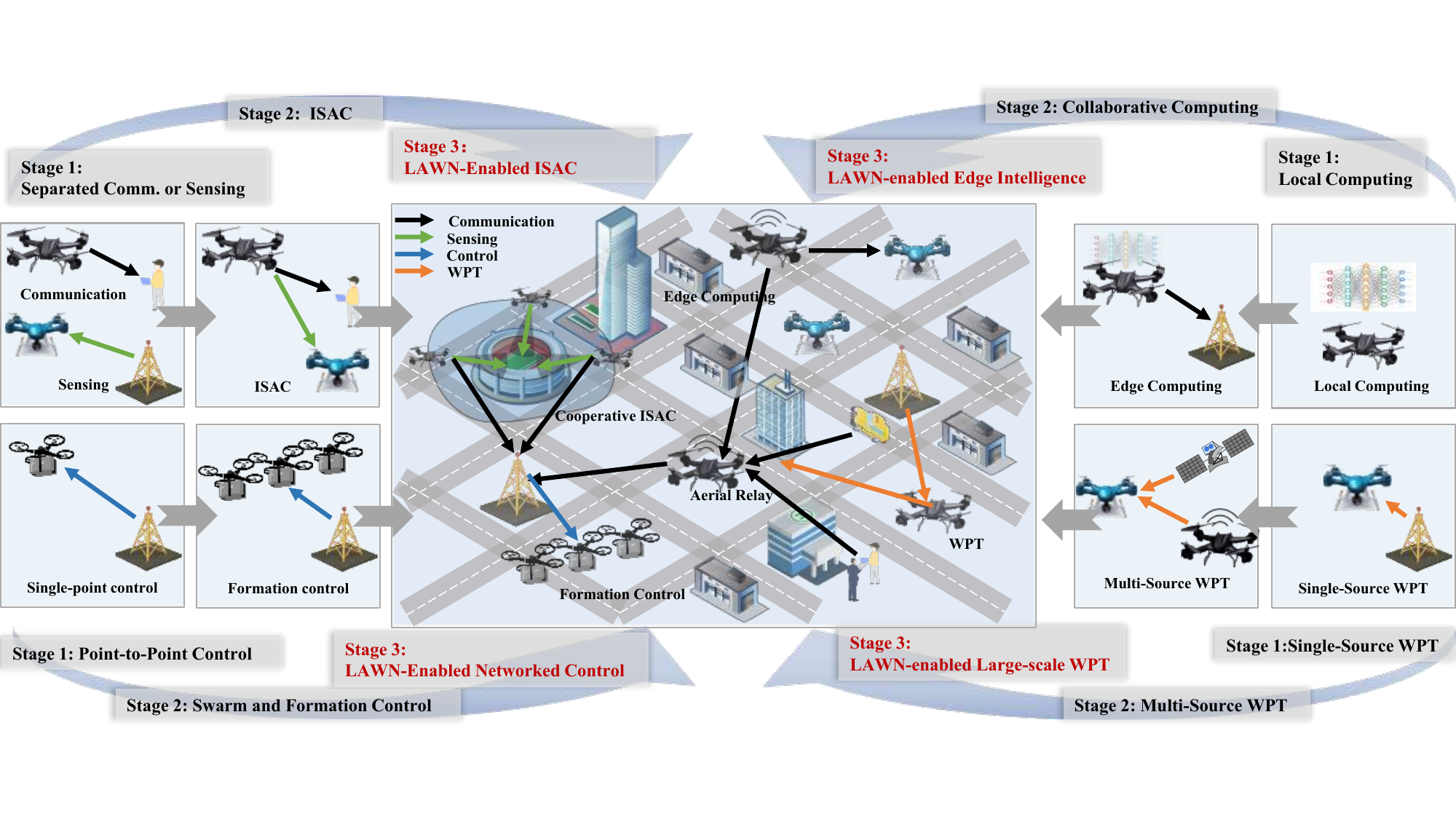}
    \caption{The evolution perspective of LAWNs, highlighting the transition from separated designs toward cooperative and ultimately intelligent integration across various functionalities.}
    \label{figfunc}
\end{figure*}

ISAC represents a transformative paradigm that unifies wireless communications and environmental sensing into a single framework \cite{liu2022integrated,mao1,mao2,wu2025toward1, mao3,mao4}. By adopting a co-design architecture, ISAC eliminates the need for deploying separate sensing or communication subsystems on aerial platforms within LAWNs. This integration not only reduces the spectral footprint, hardware redundancy, and energy consumption, but also enhances the agility of the system. For instance, a UAV transmitting communication signals to ground users can simultaneously leverage the backscattered echoes for obstacle detection or trajectory tracking of neighboring UAVs \cite{jiang2024uav}, thereby avoiding reliance on dedicated radar or LiDAR sensors.

In LAWNs, UAVs are typically deployed as distributed sensing nodes to extend coverage and provide fine-grained environmental awareness. The evolution of ISAC in such networks can be conceptually structured into three progressive stages, as illustrated in Fig.~\ref{figfunc}.
\begin{itemize}
    \item \textbf{Stage 1 (Separated systems)}: 
    In this stage, radar sensing and wireless communication are implemented as separate subsystems with independent hardware and spectrum usage. This decoupled architecture leads to spectrum congestion, hardware redundancy, and excessive energy consumption, which is particularly restrictive for UAVs with stringent payload and endurance constraints.
    \item \textbf{Stage 2 (ISAC systems)}: In this stage, each UAV integrates sensing and communication functionalities into a unified platform by sharing hardware and spectral resources. This design improves spectral efficiency, reduces payload burden, and enhances cost-effectiveness compared to separated systems. However, its benefits remain limited to individual UAVs, and scalability across large networks is not yet fully achieved.
    \item \textbf{Stage 3 (LAWN-enabled ISAC systems)}: In this stage, multiple UAVs collaborate within a LAWN to enable distributed and cooperative ISAC. By exploiting spatial diversity, synchronized waveform design, and joint signal processing, UAVs collectively form a virtual large-aperture array, thereby achieving high-resolution wide-area sensing and robust communication coverage. This networked ISAC paradigm represents a fundamental shift toward scalable, resilient, and persistent UAV operations.
\end{itemize}
This evolutionary trajectory highlights the maturation of ISAC from a dual-functional system into a network-wide intelligent infrastructure, which is particularly crucial for supporting the burgeoning LAE. In such scenarios, UAV networks serve as a vital enabler for applications ranging from urban logistics and aerial inspections to environmental monitoring and emergency response. Through cooperative sensing, these UAV networks can significantly enhance angular resolution and achieve high-fidelity environmental imaging, thereby meeting the stringent requirements of safety, reliability, and responsiveness in low-altitude economic activities.
For example, the framework in \cite{wang2025transmit} demonstrated how joint optimization can improve target detection in cluttered environments while mitigating multi-user interference, thereby boosting both sensing accuracy and communication reliability. Moreover, ISAC inherently enables implicit channel state information acquisition through environmental sensing, reducing dependence on conventional pilot-based estimation and further improving spectral efficiency \cite{pyo2025robust,10791452,10543024}. This efficiency gain not only conserves limited spectrum resources but also ensures robust connectivity for diverse services in the LAE.

In addition to its dual communication and sensing capabilities, ISAC fundamentally expands the operational capabilities of UAV networks by enabling advanced spatial awareness and autonomy, which is a key technological driver for the LAE. Applications such as high-precision three-dimensional (3D) mapping, dynamic trajectory planning, and swarm coordination become feasible when sensing and communication are tightly integrated. For instance, \cite{lu2024deep} introduced a cooperative uplink-downlink reconstruction framework, where base stations performed active sensing in the downlink while passive sensing was realized through base station-user equipment cooperation in the uplink. 
The fusion of complementary sensing results yielded higher spatio-temporal resolution, broader coverage, and improved reconstruction accuracy.
Similarly, environmental data generated via ISAC was shown to enhance real-time collision avoidance and collaborative perception in UAV swarms \cite{wang2024cooperative}, supporting safer autonomous operations in dense and dynamic LAAs. 
Advanced fusion strategies, such as the multilevel weighted fusion framework in \cite{zhang2025novel}, further improved UAV localization by combining symbol-level and data-level information, thereby enhancing estimation performance. Moreover, \cite{hu2025collaborative} proposed a collaborative ISAC-based framework for UAV-terrestrial base station cooperation, achieving high-precision localization alongside reliable communication in scenarios with multiple moving users. Plugging ISAC into LAWNs not only addresses the challenges of spectral congestion and payload constraints. It also enables distributed sensing, collaborative perception, and intelligent UAV autonomy. This progression underscores ISAC’s pivotal role in transforming LAWNs into intelligent, resilient, and spectrum-efficient infrastructures that can support mission-critical low-altitude applications such as urban monitoring, disaster response, traffic management, and environmental surveillance.

\subsection{Networked Control}
Networked control serves as a cornerstone of distributed cyber-physical systems, enabling the seamless integration of communication and control functions through shared network infrastructures \cite{meng2024communication,jin2025predictive}. By facilitating real-time exchange of sensory information and control commands among distributed nodes, it provides a scalable mechanism for coordinated operation and decision-making.

In LAWNs, networked control interconnects UAV nodes into a unified control architecture that supports collaborative swarm operation, dynamic task allocation, and efficient path optimization. As shown in Fig.~\ref{figfunc}, the progression of networked control in UAV-enabled systems can be conceptually structured into three evolutionary stages:
\begin{itemize}
    \item \textbf{Stage 1 (Point-to-point control)}: In this stage, UAVs operate independently under local autopilots or ground-based commands with minimal inter-UAV coordination. Although easy to implement, this mode lacks scalability and situational awareness, which limits its applicability in complex and large-scale missions.
    \item \textbf{Stage 2 (Swarm and formation control)}: In this stage, UAVs are interconnected through communication links to maintain predefined and relatively rigid formations for executing synchronized maneuvers and cooperative tasks such as surveillance or payload transport. Distributed control strategies and feedback mechanisms ensure stability and cohesion, but the system remains constrained to fixed team sizes and formation patterns, which limits adaptability in highly dynamic environments.
    \item \textbf{Stage 3 (LAWN-enabled networked control)}: In this stage, UAVs become integral nodes of a large-scale, self-organizing, and flexible LAWN. Unlike Stage 2, where UAVs operate within fixed formations, the UAV swarm structure in Stage 3 is highly dynamic and elastic: the number of UAVs, their roles, and the control topology can be adaptively reconfigured in real time according to mission demands. For instance, when a hotspot such as a mass gathering or disaster area emerges, additional UAVs can be rapidly mobilized to reinforce communication coverage and cooperative control. This paradigm represents a shift from static formation-based operations to resilient, scalable, and adaptive network-wide control enabled by UAV mobility and pervasive information sharing.
\end{itemize}
This evolutionary trajectory highlights how LAWNs transform rigid and isolated UAV control architectures into intelligent, resilient, and cooperative swarm systems. Such advancements are instrumental in supporting the LAE, where networked UAV swarms serve as the backbone for mission-critical LAE applications including urban surveillance, dynamic aerial coverage, and logistics distribution \cite{mozaffari2019tutorial}, all of which require precise synchronization and adaptive coordination in complex low-altitude environments. Through real-time sharing of control signals and sensory information, each UAV dynamically adjusts its trajectory and task allocation based on both local observations and global swarm feedback, thereby ensuring efficient, safe, and scalable operations across diverse LAE scenarios.

This conceptual evolution is further supported by empirical research, where LAWNs-enabled networked control has been shown to improve robustness and adaptability in real-world conditions, providing essential technical support for reliable operations in LAE scenarios. For instance, \cite{chen2025efficient} proposed a multi-UAV cooperative framework specifically designed for dynamic electromagnetic environments, leveraging UAV mobility, adaptive coverage, and reliable line-of-sight connectivity to enhance perception and overall system robustness. Complementing this approach, closed-loop feedback mechanisms were shown to significantly improve swarm adaptability under interference, uncertainty, and partial node failures \cite{qiao2022communication}. Furthermore, \cite{ping2024communication} introduced a distributed optimal control framework for multi-UAV systems that simultaneously supported formation management, obstacle avoidance, and connectivity maintenance while adaptively optimizing communication resources.

Beyond robustness and adaptability, recent studies highlighted the benefits of joint optimization across sensing, communication, and control to achieve end-to-end quality of service (QoS), a capability that can directly enhance mission reliability in critical LAE applications. For instance, \cite{wang2023qos} proposed a joint optimization framework for UAV-based positioning, where UAVs acted as aerial anchors to ensure reliable localization in challenging environments. This approach reduced service failure rates by over $75\%$ and resource consumption by $80\%$, demonstrating the efficiency of cross-domain integration. Air-ground collaboration was also explored in \cite{chen2022formation}, where a formation control strategy incorporating spatial synergy enabled UAVs to provide robust relay communication while maintaining coordinated positioning. Likewise, \cite{zhang2020unmanned} presented a lightweight, UAV-based emergency communication framework employing commercial micro-UAVs. By integrating body-conformal omnidirectional antennas with optimized formation strategies, the system enhanced cooperative transmission and ensured reliable coverage in harsh terrains such as mountainous canyons.

In summary, networked control in LAWNs has evolved from isolated point-to-point architectures to intelligent, distributed, and adaptive control infrastructures. This progression lays the foundation for future networked swarm intelligence. It is capable of operating reliably in dynamic, uncertain, and resource-constrained low-altitude environments, which are increasingly relevant for crowded urban air mobility, large-scale event management, and disaster-stricken regions.

\subsection{Mobile Edge Computing}

Mobile edge computing (MEC) seamlessly integrates computational resources into the network edge \cite{li2025edge}, enabling data processing and inference to occur locally at edge nodes such as UAVs, base stations, and edge servers. By minimizing the need to transmit large volumes of raw data to distant cloud data centers, this approach effectively mitigates latency and alleviates bandwidth congestion, thereby enhancing system efficiency, scalability, and resilience.

In the context of LAWNs, the adoption of MEC significantly elevates system performance by exploiting the distributed, mobile, and cooperative nature of UAVs. As depicted in Fig.~\ref{figfunc}, the evolution of MEC within UAV-enabled networks can be conceptualized in three progressive stages:

\begin{itemize}
\item \textbf{Stage 1 (Local computing)}: In this initial stage, each UAV independently executes computational tasks using its limited onboard processing capabilities. While this setup is suitable for lightweight tasks, the constrained computational resources and energy limitations often result in delays when executing more complex algorithms, thereby impeding real-time decision-making and autonomous operations.
\item \textbf{Stage 2 (Edge-enabled collaborative computing)}: At this stage, UAVs offload computational tasks to nearby static edge resources, such as base stations or terrestrial edge servers. This enables faster response times and more efficient resource utilization compared to local computing; however, the fixed nature of the edge nodes still limits flexibility in highly dynamic environments.
\item \textbf{Stage 3 (LAWN-enabled MEC)}: In this advanced stage, UAVs evolve into dynamic, mobile computing nodes that collectively form a self-organizing aerial-ground edge network. Unlike Stage 2, where edge resources are fixed, UAVs in this stage are capable of mobilizing and coordinating as needed, dynamically deploying additional communication and computational capacity in response to hotspots or mission-critical requirements. Computation is performed adaptively through federated learning, proactive task offloading, and mobility-aware resource orchestration. This stage marks a paradigm shift from static edge collaboration to a flexible, network-wide computing infrastructure empowered by UAV mobility.
\end{itemize}

This progression is especially pivotal for enabling and sustaining LAE activities, where UAV operations must continuously adapt to dynamic and uncertain environments. LAE-relevant tasks, such as real-time logistics routing, aerial infrastructure inspection, target tracking, and environmental perception, require low-latency and highly reliable computation. Traditional cloud-centric computing models are often constrained by high latency due to long-distance backhaul links, which can hinder time-sensitive applications. In contrast, MEC enables lightweight computational models to be deployed directly on UAVs or adjacent edge nodes, facilitating immediate decision-making and significantly enhancing responsiveness, autonomy, and mission reliability \cite{huynh2022uav}. This results in more efficient and reliable execution across a diverse array of low-altitude economic scenarios.

Furthermore, the distributed nature of LAWNs, with multiple UAVs operating cooperatively, lends itself well to distributed computing paradigms such as federated learning. These paradigms offer scalable, privacy-preserving, and resource-efficient solutions for LAE operations. By training models collaboratively across UAV nodes without sharing raw data, federated learning not only ensures privacy but also reduces communication overhead \cite{zhang2025latency}. For example, \cite{deng2022uav} introduced a UAV-enabled MEC framework for air-ground integrated networks, where UAVs acted as aerial base stations to provide adaptive coverage and edge computing support for Internet-of-Things (IoT) devices under stringent energy and accuracy constraints. Similarly, \cite{wang2023uav} proposed a two-tier hierarchical federated learning framework, where UAVs served as mobile relays, enhancing connectivity to a central server while simultaneously conducting local computations, thus improving scalability in large-scale, infrastructure-limited environments.

Beyond distributed learning, recent advancements in UAV-enabled MEC have explored a variety of application domains and performance metrics directly applicable to LAE services. For instance, \cite{zhang2025age} introduced a UAV-assisted IoT data collection framework under segmented fading channels, which achieved reduced AoI, lower energy consumption, and faster mission completion times through collaborative UAV operations. In \cite{shaikh2025flying}, a flying-edge computing framework for smart agriculture was proposed, where UAVs equipped with AI-powered sensors perform tasks such as disease detection and irrigation control on-board, enabling real-time decision-making in areas with limited connectivity. Additionally, \cite{zhang2024minimizing} presented a UAV-assisted MEC framework that optimizes UAV deployment and computation offloading to minimize response delays, ensuring rapid adaptability to dynamic operational environments.

\subsection{Wireless Power Transfer}
Wireless power transfer (WPT) delivers energy to electronic devices through radio-frequency electromagnetic waves or inductive coupling, thereby enabling continuous power replenishment without physical connections \cite{zhang2018wireless}. In LAWNs, WPT provides an effective solution to the endurance limitations of battery-powered UAVs, extending mission duration, reducing task interruptions, and minimizing maintenance costs \cite{xu2018uav}. Its integration with UAV swarms further enhances operational sustainability, making long-duration and high-frequency missions feasible.
The evolution of WPT in UAV-enabled networks can be conceptually divided into three progressive stages, as illustrated in Fig.~\ref{figfunc}.
\begin{itemize}
    \item \textbf{Stage 1 (Single-source WPT)}: In this stage, energy is wirelessly transmitted from a single ground-based source (e.g., base station, charging pad, or access point) to UAVs within short line-of-sight range. Typical technologies include near-field inductive coupling and low-power radio-frequency beaming, which are suitable for safe and localized recharging. However, the energy transfer is constrained by limited coverage, rapid decay over distance, and insufficient support for mobile UAV swarms.
    \item \textbf{Stage 2 (Multi-source WPT)}: In this stage, multiple heterogeneous energy transmitters jointly power UAVs, forming a multi-point-to-multi-point architecture. Energy sources may include ground stations, high-altitude platforms (e.g., airships, balloons), satellites, and cooperative aerial relays. This architecture significantly extends spatial coverage, enhances continuity of energy supply, and allows coordinated energy delivery across terrestrial, aerial, and spaceborne infrastructures. Nevertheless, the supply pattern is still largely infrastructure-driven and limited in flexibility.
    \item \textbf{Stage 3 (LAWN-enabled WPT network)}: In this stage, UAVs themselves evolve into active energy nodes within a self-organizing WPT network. Beyond being consumers, UAVs can dynamically act as transmitters, relays, or cooperative chargers, redistributing harvested or stored energy across the swarm. By integrating WPT with trajectory design, energy-aware scheduling, and distributed control, LAWNs enable an elastic and resilient power-sharing infrastructure. This marks a paradigm shift from static multi-source provisioning to a fully networked, scalable, and self-sustaining WPT system, supporting persistent and large-scale UAV operations.
\end{itemize}
Building on this evolution, cooperative energy sharing among UAVs emerges as a critical enabler of swarm endurance, which is essential for sustaining continuous operations in the LAE. UAVs equipped with wireless charging modules can serve as aerial charging stations, replenishing the batteries of neighboring UAVs with depleted energy. Coordinated energy scheduling and joint task allocation allow UAV swarms to dynamically balance energy distribution while maintaining mission performance, thereby ensuring uninterrupted execution of LAE-related missions. This mechanism is particularly advantageous for large-scale sensing, urban surveillance, infrastructure inspection, or logistics services within the LAE, where UAVs must operate over extended low-altitude areas without frequent returns to ground stations, thus improving operational efficiency and economic viability.

Extending the concept of cooperative energy sharing, recent research has explored the implementation of LAWNs-enabled large-scale WPT in practical scenarios, which can serve as a key technological enabler for various LAE applications. In \cite{hu2020wireless}, a cooperative wireless-powered MEC framework was proposed, where a grid-powered access point provided both computational services and laser-based charging to a UAV. The UAV, in turn, functioned as an information relay, energy relay, and mobile edge server. Similarly, \cite{yin2019uav} introduced a UAV-assisted cooperative system that employed simultaneous wireless information and power transfer (SWIPT), enabling UAVs to act as mobile relays powered directly by RF energy. Beyond cooperative relaying, heterogeneous architectures were explored. For instance, \cite{liu2024joint} proposed a UAV-enabled MEC framework where UAVs served as both wireless chargers and mobile computing enhancers for energy-constrained IoT devices. Likewise, \cite{xie2018throughput} considered a harvest-then-transmit protocol in UAV-enabled wireless powered communication networks, while \cite{wang2025practical} investigated trajectory optimization in obstacle-rich environments to maximize energy delivery efficiency. At the ground-user level, \cite{li2023energy} optimized 3D UAV trajectories for fair and efficient power allocation among distributed nodes, and \cite{shi2023two} addressed charging of energy-constrained sensor nodes with unknown locations via UAV-based WPT strategies. In summary, WPT in LAWNs evolves from static ground-based charging to a dynamic, network-wide energy sharing paradigm. By leveraging UAV mobility, swarm cooperation, and joint optimization with sensing and communication, LAWNs enable persistent, autonomous, and large-scale aerial operations. When combined with ISAC and edge intelligence, WPT does not merely extend flight time. It provides the energy backbone that sustains intelligent perception, resilient communication, and adaptive networked control for low-altitude economic scenarios such as long-endurance surveillance, infrastructure inspection, cargo transport, and emergency energy replenishment. 

{From a system perspective, multi-functional LAWN designs are best understood through task integration rather than as a collection of isolated functions. A concrete mission may simultaneously impose requirements on sensing accuracy, communication reliability, computation offloading, control stability, and energy sustainability. Effective LAWN operation accordingly relies on a coordinated mechanism that maps mission objectives into coupled decisions on beamforming, scheduling, computation placement, controller updates, and power management across heterogeneous aerial and ground nodes. For instance, ISAC determines the timeliness and quality of state and environment information available to edge intelligence and networked control, while control and trajectory decisions in turn reshape channel conditions, interference patterns, and energy consumption. Computation offloading decides how quickly complex functions such as online planning and collaborative perception can be updated and executed in a closed loop, subject to latency and reliability constraints imposed by the communication layer. Meanwhile, WPT further constrains the long-term energy budget of aerial platforms and creates trade-offs between aggressive functional performance and sustainable operation. In practice, multi-functional LAWN design must therefore balance the competing demands of ISAC, edge intelligence, networked control, and WPT within a unified task-centric framework, rather than optimizing each function independently.}

\section{Multi-Level Defenses of LAWNs}\label{sec5}

\subsection{Privacy preservation}

Privacy preservation of LAWNs refers to protecting sensitive information and preventing unauthorized access or disclosure of data related to the operation, environment, and participants within LAWNs \cite{cordill2025comprehensive,kaleem2025quantum}. 
By addressing the privacy concerns of the above issues, LAWNs can foster user trust, enable wider adoption, and comply with legal and ethical requirements, ultimately unlocking the full potential of LAWN technology while minimizing the risks associated with data breaches and misuse. Currently, researchers have adopted differential privacy, federated learning, homomorphic encryption, secure multi-party computation, destination obfuscation, game theory and blockchain-based techniques to realize the privacy preservation of LAWNs on sensing, communication, computing, and control aspects.

\begin{figure}[tb]
  \centering
  \includegraphics[width=8.5cm]{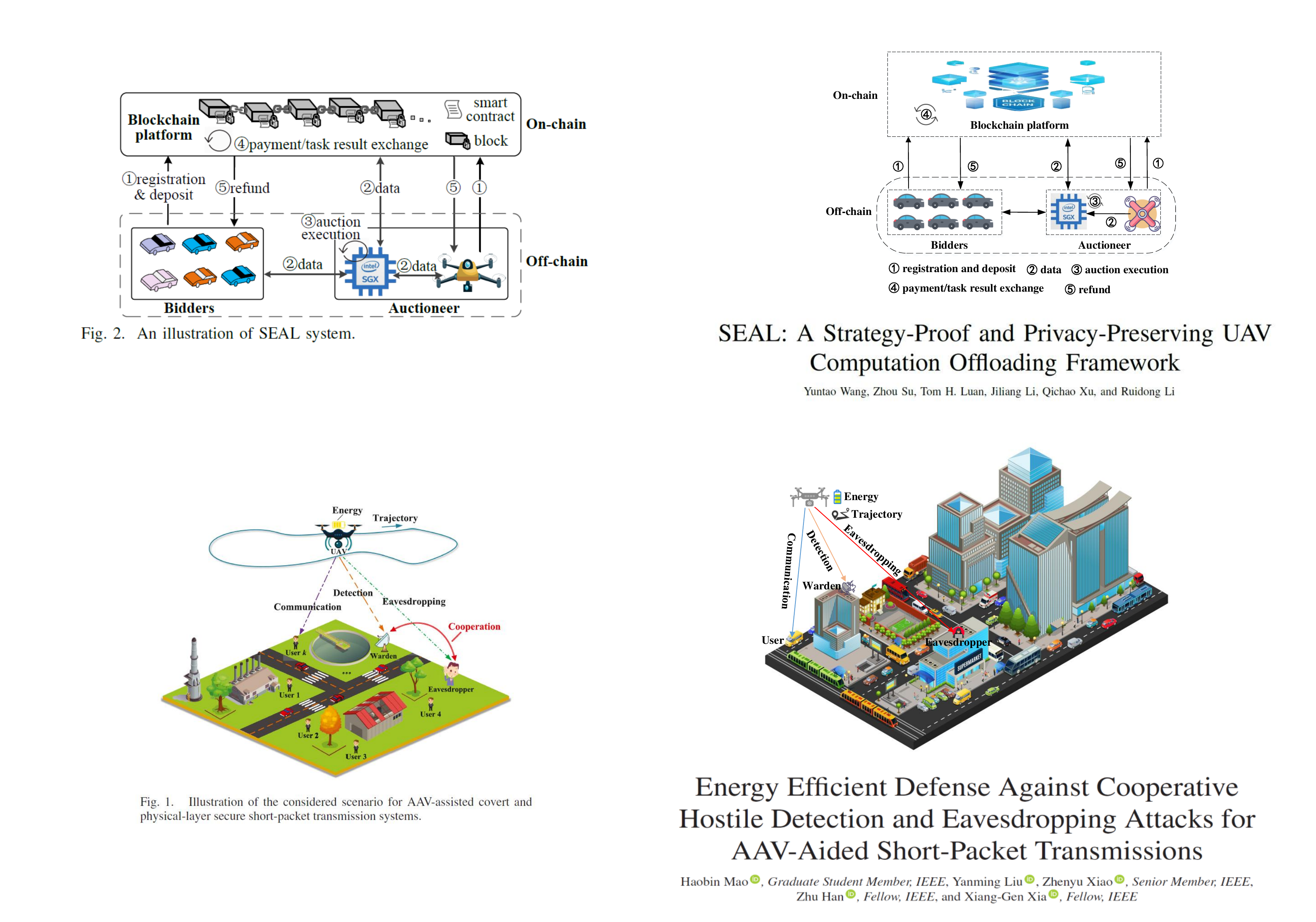}\\
  \caption{The framework of the proposed SEAL defense scheme in UAV offloading scenario \cite{wang2023seal}.}\vspace{-0.6cm}
  \label{2}
\end{figure}

For example, for privacy preservation on the sensing aspect, to ensure privacy in UAV-based delivery and sensing, \cite{liu2023decentralized} introduced a game-theoretic routing framework that aims to protect UAV trajectory privacy. The approach leverages a task-time graph and a non-cooperative potential game to optimize sensing, routing, and task selection. This framework demonstrates effective privacy preservation while ensuring timely delivery and achieving efficiency.
In terms of communication privacy, \cite{chen2020traceable} presented a traceable and privacy-preserving authentication scheme for UAV applications, employing digital signatures to ensure integrity, confidentiality, and manage sensitive area access. 
Addressing security in UAV surveillance, \cite{deebak2020smart} introduced a smart Internet of drone (S-IoD) framework with IPA and a lightweight privacy-preserving scheme (L-PPS). This L-PPS expedites authentication between devices through a secret token and dynamic user authentication, mitigating computational overhead.
For privacy in UAV-assisted IoT computation offloading, \cite{wei2021uav} identified that current deep reinforcement learning-based methods leak UAV offloading preferences. The authors introduced a privacy-preserving method that integrates differential privacy into deep reinforcement learning to protect these preferences, including formal analysis of security and utility.
Similarly, as shown in Fig. \ref{2}, \cite{wang2023seal} addressed the critical requirements of incentives, privacy, and fairness when offloading UAV computation tasks to ground vehicles. Their proposed framework overcomes issues of manipulation and privacy leakage by using a strategy-proof reverse combinatorial auction. In addition, fairness is managed through smart contracts and hash-chain micropayments, complemented by a privacy-preserving off-chain auction protocol utilizing a trusted processor.
Focusing on control-related privacy, \cite{pan2019unmanned} addressed the privacy-preserving navigation (PPN) problem for visual-based UAVs, aiming to minimize privacy violations within specific sub-areas while covering a target region through a heuristic path planning solution. The work of \cite{liu2024privacy} tackled privacy concerns in cooperative UAV operations by proposing a decentralized framework that preserves trajectory and destination privacy through probabilistic destination obfuscation and a noncooperative Bayesian game for joint routing and charging.

\subsection{Physical-Layer security}

Physical layer security of LAWNs refers to protecting the confidentiality, integrity, and availability of information transmitted and processed within LAWNs by exploiting the inherent characteristics of the physical layer itself \cite{wang2021physical,10168298}. This protection is achieved without relying solely on traditional cryptographic methods. By addressing the physical layer security concerns of the above issues, LAWNs can enhance resilience against sophisticated attacks, enable robust operation in contested environments, and provide lightweight security solutions suitable for resource-constrained UAV platforms, ultimately contributing to safer and more reliable UAV deployments. In general, physical layer security is commonly classified into two fundamental paradigms, namely the wiretap channel and covert communication. Currently, researchers have adopted beamforming techniques, artificial noise injection, cooperative jamming, power allocation strategies, and intelligent reflecting surfaces (IRS) to realize the physical layer security of LAWNs on sensing, communication, computing, and control aspects. 
For example, for physical layer security in integrated sensing and communication, \cite{xiu2024improving} explored a system where a UAV transmits ISAC waveforms to securely communicate with IoT devices while simultaneously sensing the environment. This approach optimizes UAV trajectory and beamforming to maximize the average communication secrecy rate while ensuring accurate sensing capabilities. In the context of covert ISAC in LAWN systems, \cite{xx} studied covert ISAC for multi-antenna UAV systems using a unified waveform for communication and sensing, revealing a trade-off between covert communication and sensing accuracy by exploiting prior information of the adversary.
Regarding physical layer security in UAV communication, \cite{li2021physical} addressed wire-tap channel communications in 5G networks with eavesdroppers by proposing a virtual antenna array with collaborative beamforming. To enhance security, UAV positions and excitation current weights are optimized, aiming to improve total secrecy rates and reduce sidelobe levels through an improved multi-objective dragonfly algorithm. 
Furthermore, to satisfy the heterogeneous security requirements in low-altitude UAV systems, where some users demand secrecy while others require covertness, the authors in \cite{10681761} investigated multi-user cooperative secret and covert communications for UAV networks.
For the physical layer security aspect within computing, \cite{zhou2019secure} investigated a secure UAV MEC system where ground users offload tasks to a legitimate UAV amidst eavesdropping UAVs. It enhances security by employing jamming signals from the legitimate UAV and users to maximize minimum secrecy capacity, jointly optimizing aspects like offloading ratio and computing capacity to balance security and latency. \cite{nasir2024secure} focused on secure and energy-efficient UAV-MEC with RIS assistance, aiming to maximize minimum energy efficiency against an eavesdropper. The research ensured security through artificial noise transmission and optimizes computation tasks/CPU frequency allocation alongside other parameters, demonstrating strong performance even with an eavesdropper present.
Focusing on physical layer security in control, \cite{xu2025collaborative} proposed a secret and covert transmission strategy for ground-to-UAV links using non-orthogonal multiple access (NOMA). This method allows simultaneous transmissions with different security needs, enabling an adversary to attempt both data eavesdropping and transmission detection. Security is rigorously ensured with optimization targeting maximized rates through beamforming and trajectory control. \cite{chen2021uav} improved covert communication range by using a UAV relay and finite blocklength transmissions, aiming to maximize secure bits against a warden. It achieves this by optimizing the warden’s threat potential and jointly controlling transmission parameters under error constraints, leveraging signal uncertainty at the warden for physical layer covertness.

\subsection{Network-Layer security }

Network layer security of LAWNs refers to protecting the security of data as it is routed and managed across the network infrastructure of LAWNs. 
This involves securing the routing protocols, preventing unauthorized access, and defending against various network-based attacks that target the data traversing the network. This protection is vital for LAWNs due to their reliance on wireless communications and distributed architecture. 
By addressing the network layer security concerns of the above issues, LAWNs can ensure reliable data delivery, prevent unauthorized access, mitigate the impact of network-based attacks, and guarantee the integrity of the network infrastructure. This leads to more trustworthy and secure UAV operations, especially in sensitive applications like surveillance, delivery, and infrastructure inspection.
Currently, researchers have adopted intrusion detection systems, secure routing protocols, authentication and access control mechanisms, and blockchain technologies to realize the network layer security of LAWN on sensing, communication, computing, and control aspects.

For example, addressing network layer security challenges for UAV-based sensing, \cite{sedjelmaci2017hierarchical} designed an intrusion detection and response system to identify and mitigate attacks that compromise the integrity of sensed data, including false information dissemination and GPS spoofing. 
To ensure network layer security in UAV communication, \cite{fotohi2022self} introduced self-adaptive intrusion detection (SID)-UAV, which employs a self-matching system to detect safe communication routes, coupled with a distributed architecture for investigation and response to malicious UAVs. 
The work of \cite{mao2024energy} enhanced network layer security for autonomous aerial vehicle (AAV)-aided short-packet transmissions against detection and eavesdropping by optimizing for maximum covert secrecy energy efficiency through a dual-loop iterative algorithm for user association, power control, and trajectory planning.
Addressing network layer security in UAV edge computing, \cite{sedjelmaci2019efficient} presented a cyber-defense mechanism based on a non-cooperative game to protect against network and offloading threats. The solution aims to maximize protection while minimizing the computational and energy costs associated with the defense. In response to the data leakage risk in UAV-assisted traffic surveillance at the network layer, \cite{garg2018uav} utilized a probabilistic data structure approach for cyber-threat detection and efficient data management, minimizing computational overhead in the process.
Regarding network layer security for control, \cite{ale2024advanced} proposed using sliding mode control to enhance UAV security. The system dynamically reconfigures to adapt to network-level threats like GPS spoofing and jamming, ensuring robust control even under attack through anomaly detection and fault tolerance. \cite{xie2025blockchain} tackled network layer access control challenges in distributed UAV swarm control by using a blockchain-based scheme. It ensures secure authentication and data integrity against identity spoofing and data injection, employing dynamic key management for task and identity groups.



\section{Advanced AI-driven LAWNs}\label{sec6}
Traditional model-based optimization and control frameworks, while effective under static and predictable conditions, struggle to cope with the high-dimensional, dynamic, and uncertain nature of low-altitude environments. In this section, we delve into the integration of advanced AI technologies into LAWNs, focusing on the deployment of computationally efficient large AI models and agentic AI. In particular, large AI models enhance cognitive capabilities for complex decision-making and environmental understanding, and agentic AI empowers LAWNs to be autonomous and collaborate in dynamic environments, improving the scalability and resilience.
\subsection{Large AI Model-based LAWNs}
Large AI models, characterized by their extensive parameter dimensions and training on massive datasets, have made notable advancements across various domains, including natural language processing, computer vision, and autonomous systems. Prominent models, e.g., generative pre-trained transformers (GPT), bidirectional encoder representations from transformers (BERT), and similar architectures, possess the ability to learn intricate patterns and generalize across a wide range of tasks. This makes them particularly suitable for dynamic and complex environments. Hence, the integration of large AI models into LAWNs holds substantial potential for enhancing the cognitive capabilities of UAVs, facilitating real-time decision-making in unpredictable environments. A common approach to this integration involves utilizing large language models (LLMs) in physical layer designs, where pre-trained models are deployed at both the UAVs and controller terminal to enable effective online inference. For example, the work of \cite{Xiao2024} illustrates how large AI models can be employed in UAV path planning, where they enable the UAVs to dynamically adjust their flight paths based on real-time environmental data, such as wind and obstacle conditions. Similarly, the work of \cite{Javaid2024} explores how large language models (LLMs) can enhance UAV communication by enabling more intuitive interactions between UAVs and ground control stations. This capability allows for more adaptive and flexible communication, improving operational coordination, especially in complex scenarios like disaster relief or surveillance operations.

Despite their impressive capabilities, deploying large AI models in LAWNs presents significant challenges, primarily due to the substantial computational and memory requirements of these models. The increasing size of model parameters, often reaching billions, imposes a heavy computational burden, which exceeds the resources available in many UAVs and edge devices. The need for real-time decision-making further complicates the deployment of such models, as low latency is critical in many applications. To circumvent these challenges, two prominent distributed strategies have been proposed, i.e., \textit{model distribution} and \textit{data distribution}. Both approaches aim to alleviate the computational load on individual UAVs and edge devices by distributing the processing requirements across a network of devices.

Model distribution involves partitioning a large AI model across multiple devices or computational nodes, such as UAVs or edge servers. Instead of running the entire model on a single device, each UAV or edge server processes a portion of the model. This distributed approach allows for parallel processing, which significantly reduces the computational burden on each individual device. The work of \cite{Piggott2023} demonstrated how model distribution can be effectively implemented in UAV networks, where large AI models are divided across several UAVs, enabling real-time inference without overwhelming the resources of any single UAV. This technique enhances the scalability of large models, making it possible to deploy them in a distributed manner without compromising system performance. Data distribution, on the other hand, focuses on distributing the data processing tasks across multiple devices to minimize the need for large-scale data transmission. In the context of LAWNs, this approach ensures that data collected by UAVs is processed locally, reducing the communication overhead and enabling real-time decision-making. The work of \cite{Javaid2024} highlighted the use of federated learning in UAV networks, a data distribution method that allows multiple UAVs to collaboratively train a shared model by exchanging model updates rather than raw data. This approach not only preserves data privacy but also reduces the bandwidth required for communication between UAVs and ground control systems, making it particularly beneficial in applications where security and privacy are paramount.

However, it is important to note that both model distribution and data distribution entail substantial resource-intensive fine-tuning processes \cite{10415235}. Therefore, high-dimensional computing (HDC) has attracted increasing attention as a computationally efficient alternative \cite{chang2023recent}. HDC is a brain-inspired computing paradigm in which information is encoded into extremely high-dimensional vectors, often containing tens of thousands of components. Such representations naturally support large-scale parallelism and enable rapid processing, while exhibiting strong robustness against noise, distortions, and missing data. 
A fundamental property of these high-dimensional vectors is that independently generated vectors are almost orthogonal to one another. This behavior results from the randomization procedures during vector construction, which ensure that distinct vectors share very limited overlap \cite{basaklar2021hypervector}. Consequently, meaningful representations tend to have large Hamming distances and low cosine similarity, providing substantial capacity for encoding diverse information with minimal interference. These characteristics enable HDC to rely on simple arithmetic operations, including addition for combining multiple vectors and multiplication for binding variables with their associated values \cite{kanerva2009hyperdimensional}. Similarity measures, such as Hamming distance and cosine similarity, are used to determine how closely two vectors align, supporting accurate retrieval and comparison. Since these operations do not require heavy matrix computations, they are highly efficient and particularly suitable for UAVs operating in LAWNs, where memory, computation capability, and energy supply are inherently limited.

HDC has also shown strong potential for improving wireless communication reliability in scenarios relevant to LAWNs. Its high-dimensional representations can tolerate interference and signal distortions, allowing communication to remain stable even when multiple devices transmit simultaneously. A representative example is the direct transmission of high-dimensional vectors from a group of sensing devices. Instead of coordinating access or assigning dedicated channels, each device simply sends the vector that corresponds to its measurement. The receiver then observes the superimposed vector and determines the underlying sensing events by comparing the received vector with a set of reference patterns. By doing so, HDC enables efficient UAV reporting without traditional access control. For example,  the authors in \cite{jakimovski2012collective} considered a scenario where the receiver detects temperature conditions and determines how many devices experienced a particular temperature by analyzing the combined vector. Furthermore, the study in \cite{kleyko2012dependable} generated composite vectors through binding and superposition to encode local sensing information at each device. The receiver successfully interpreted the embedded information using predefined base vectors, demonstrating that HDC can maintain reliable information extraction even in dense and interference-prone environments. These results indicate that HDC offers a practical and lightweight computational method for LAWNs, enabling robust communication and efficient information exchange while accommodating the limitations of UAV platforms.

\subsection{Agentic AI-based LAWNs}

Although lightweight large AI models substantially strengthen LAWNs by offering advanced perception and inference capabilities, the operational role of these models in current designs remains largely passive. In most architectures, large models are tied to predefined objectives, rely on offline pretraining, and function primarily as static function approximators that map inputs to outputs under fixed task specifications\cite{Javaid2024}. The behavior is effectively constrained by the training data distribution and objective design, which limits the ability to autonomously revise goals, reinterpret evolving mission contexts, or sustain long-horizon coordination with other agents. Such rigidity is particularly problematic in LAWNs, where UAVs operate in nonstationary environments, face time-varying and sometimes conflicting mission objectives, and must continuously interact with other aircraft and ground infrastructure.

Agentic AI offers a promising paradigm to overcome these limitations by elevating LAWNs from passive optimization engines to active, goal-directed autonomous systems. By embedding reasoning, memory, tool utilization, and self-evolution into network entities, agentic AI enables UAVs and ground infrastructure to adapt trajectories, allocate resources, and collaborate in swarms in a mission-aligned manner \cite{sapkota2025uavs,gao2025agentic}. Through continual learning and shared intent representations, agentic LAWN systems are capable of mitigating performance drift induced by mobility and partial observability, thereby sustaining resilient closed-loop operation in complex low-altitude scenarios \cite{acharya2025agentic}. In this context, large models and learning modules are no longer treated as static components; instead, they form the core of deliberative agents that can interpret goals, plan actions, and revise strategies online as the environment and mission requirements evolve.

It is worth mentioning that, within the agentic paradigm, task integration emerges as the central organizing principle for AI-based LAWNs. Rather than treating sensing, communication, control, and cognition as loosely coupled functions, the network is structured around mission-level tasks that are decomposed into several interdependent subtasks, which are allocated to heterogeneous agents and continuously re-evaluated as environments change. In particular, a task-oriented cognitive layer can be adopted to interpret high-level intent and fuse multi-modal observations, while lower layers implement coordinated functions, e.g., ISAC, wireless networked control, and edge intelligence. By doing so, UAVs and ground nodes can reason jointly about trade-offs in communication reliability, control precision, and sensing accuracy to adapt behavior based on both local context and global performance metrics \cite{feng2025multi,zhou2024comprehensive}. In this way, the mission goal is translated into consistent cross-layer decisions, followed by gradually reshaping task priorities and resource allocations based on real-time feedback from execution.

Building on this integrated view, some recent research has begun to explore agentic optimization solutions that couple mathematical coordination frameworks with autonomous decision-making. Optimization-based approaches, originally formulated in a centralized manner, are extended to distributed and agent-based schemes in which each UAV acts as a decision-making agent handling local objectives and constraints while obeying global mission requirements. For example, distributed optimization has been employed to jointly design multi-role UAV functionality switching and trajectories for secure task offloading in UAV-assisted mobile edge computing, where local agents iteratively coordinate to satisfy communication and computation constraints under dynamic conditions \cite{zhong2024distributed}. In parallel, LAWN-wide closed-loop designs have been formulated that integrate sensing, communication, and control into unified optimization problems, demonstrating that jointly tuned resource allocation and controller design can significantly enhance positioning accuracy and service reliability compared with decoupled baselines \cite{wang2024semantic,zhou2024comprehensive}. 
Complementing optimization-driven coordination, agentic AI solutions build learning-based autonomy on top of this substrate. At the single-agent level, reinforcement learning (RL) has been widely adopted to endow individual UAVs with adaptive decision-making capabilities in complex, uncertain environments. Typical implementations rely on actor–critic architectures, deep Q-networks, or proximal policy optimization to learn control and path-planning policies directly from interaction with the environment \cite{kaelbling1996reinforcement,konda1999actor,schulman2017proximal}. In soft actor–critic–based designs, for instance, a UAV agent observes its flight state and local environment and outputs continuous control actions that track reference trajectories while avoiding obstacles and following energy and communication constraints \cite{cheng2020autonomous}. The entropy-regularized objective promotes a balance between exploitation of known good behaviors and exploration of new strategies, improving robustness to modeling errors and unforeseen disturbances. From the LAWN perspective, such a controller functions as a local decision engine that interprets mission-level commands and translates them into fine actuation in real time.

When multiple UAVs and ground nodes need to coordinate, multi-agent reinforcement learning (MARL) extends this paradigm to networked settings, where centralized training and decentralized execution are applied\cite{kopic2024collaborative}. Specifically, each UAV is modeled as an agent with partial local observations and a policy that must adapt under the nonstationary dynamics induced by other learning agents. To name a few, the authors in \cite{cui2019multi} investigated MARL-based resource allocation schemes that enable UAVs to share spectral and power resources while jointly maximizing network throughput and fairness. Furthermore, the work of \cite{liang2023multi} developed a multi-UAV collaborative search and attack framework that formulated exploration, target engagement, and collision avoidance in unknown environments as a multi-agent decision problem. In fact, such learned coordination strategies are intractable to engineer manually, highlighting the value of agentic learning in realizing cooperative, mission-aware behaviors in LAWNs, even though issues such as nonstationarity, credit assignment, and training stability remain challenging.

A further step toward cognitively richer agentic AI integrates LLMs as reasoning engines for UAV agents. For example, LLM-based UAV path-planning frameworks have been proposed for industrial scenarios, where an LLM is employed to interpret mission descriptions, query external tools or simulators, and generate high-level plans that are subsequently translated into executable trajectories for individual UAVs \cite{wang2024survey}. In addition, the studies in \cite{Javaid2024,Li2025} provided a systematic overview of how LLMs can support intent understanding, dialogue-based supervision, and explainable decision-making in LAWN systems. Operational prototypes such as Net-GPT further demonstrated LLM-empowered intermediaries that maintain contextual information, enforce safety policies, and coordinate interactions among UAVs, human operators, and external services in real time \cite{Piggott2023}. Collectively, these developments point to a transition toward cognitive agentic LAWNs, where LLMs handle high-level reasoning, task decomposition, and tool orchestration, while specialized learning or optimization modules are responsible for low-level control and communication under explicit safety and regulatory constraints.

\section{LAWN implementation and management}\label{sec7}
In this section, we introduce the LAA structuring and ATM mechanisms that support the practical realization of large-scale LAWNs.

\subsection{LAA Structuring}
\begin{figure*}
    \centering
    \includegraphics[width=\linewidth]{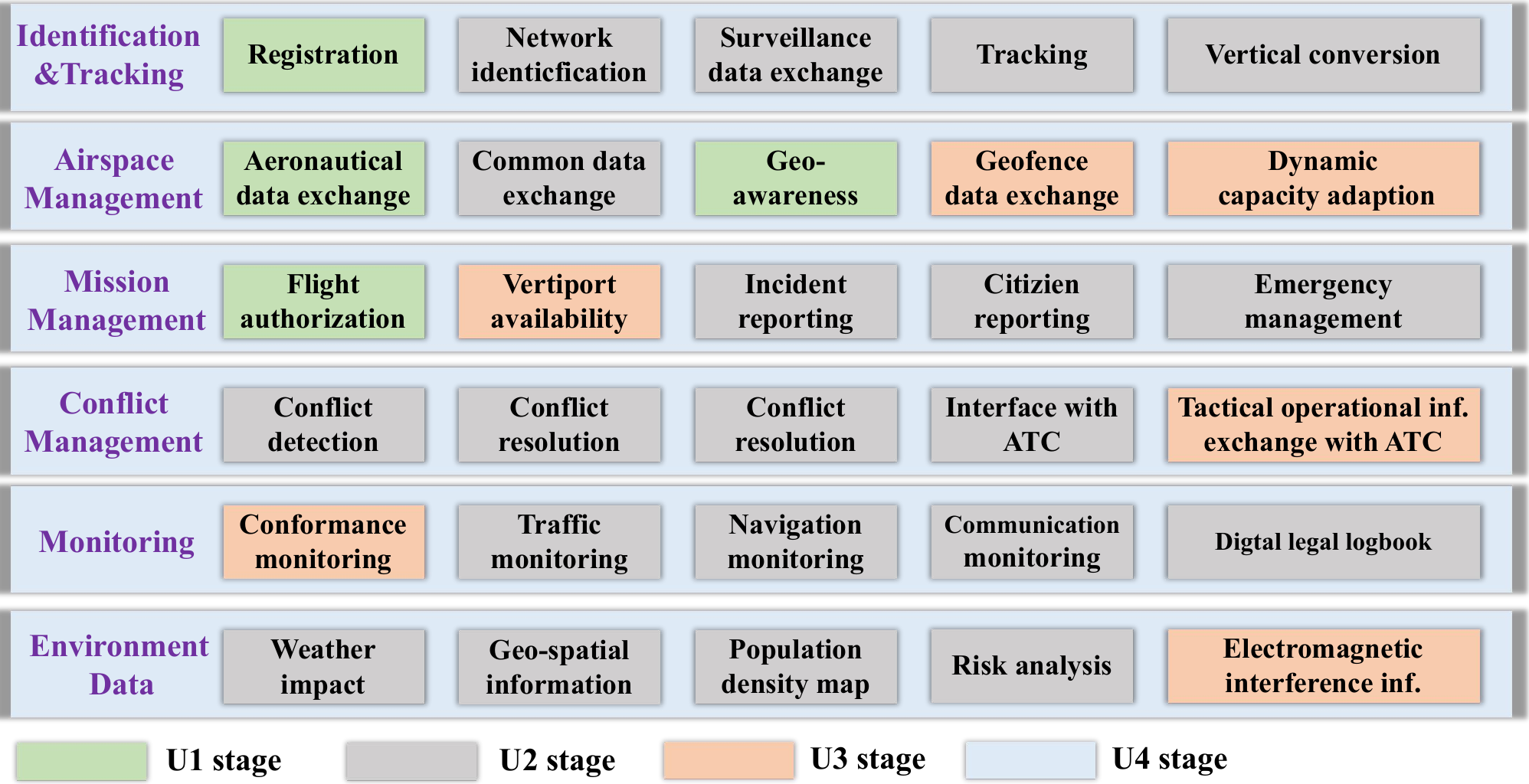}
    \caption{The supported various LAWN services associated with different U-space stages based on \cite{juntong2025low}.}
    \label{fig:placeholde123r}
\end{figure*}
LAA is a fundamental element for the successful implementation of LAWNs, where the proper organization and structuring of airspace become crucial to ensure both operational efficiency and safety\cite{jia2025hierarchical,11159633}. Nevertheless, the corresponding configuration and structuring of LAA remain in the early stage. The performance and management efficiency of LAA will directly affect the operation of urban air mobility (UAM) systems, such that the LAA structure must not only accommodate uncrewed drones but also facilitate the integration with manned aircraft.  According to the ICAO framework \cite{FAA2022}, the entire airspace is divided into categories A, B, C, D, E, and G, with the G category being non-controlled airspace, while the others are controlled airspace. LAA primarily corresponds to categories G and E, but it may partially overlap with other categories in emergencies. The ultimate goal of LAA planning is to allow for the joint operation of low-altitude aircraft and traditional aviation.
{
 Against this background, the SESAR program \cite{SESAR2017} defined a four-stage (U1-U4) development strategy for LAA, which is particularized as:
\begin{itemize}
  \item \textbf{U1}  
  corresponds to the initial stage of U-space deployment, defined in 2019 for class G airspace below 400 ft and aimed at supporting mixed crewed and uncrewed operations under relatively low traffic density. At this level, operations are typically confined to restricted or predesignated areas, and the service set remains intentionally limited. The key functions consist of basic registration of unmanned aircraft and geo-awareness capabilities, enabling operators to understand and respect airspace restrictions while laying the groundwork for more advanced U-space services in subsequent phases.
  \item \textbf{U2} builds on this foundation by enabling beyond visual LoS (BVLoS) operations in higher-density environments, extending into portions of controlled airspace. In this phase, U-space introduces a procedural interface with conventional air traffic control (ATC), allowing uncrewed operations to be coordinated with existing ATM procedures. The service portfolio expands to drone operation management functions, including flight planning and approval, real-time tracking of flights, and the provision of dynamic airspace information, thereby supporting safer and more predictable BVLoS activities.
  \item \textbf{U3} represents a further step toward advanced U-space services, targeting BVLoS operations in dense and complex airspace such as suburban areas and emerging urban air mobility corridors. This stage assumes more sophisticated and higher levels of automation in the interaction between U-space and ATC. Service capabilities are enhanced to include advanced geo-awareness, dynamic geofencing, dynamic capacity management, and conflict detection and resolution functions. According to current planning, U3-level services are expected to reach operational maturity around 2027. 
  \item \textbf{U4}  
  envisions comprehensive integration of unmanned aircraft into very dense urban airspace, with a correspondingly high level of automation for both vehicles and the supporting network and U-space infrastructure. At this stage, U-space and traditional ATM services are expected to be fully implemented and tightly coupled, enabling seamless management of crewed and uncrewed traffic in highly demanding environments. Full realization of U4 capabilities is anticipated around 2035 or later.
\end{itemize}
Along with this evolution,  the service sets associated with U1 through U4 are summarized in Fig.  \ref{fig:placeholde123r}.}
In contrast to high-altitude airspace, LAA is significantly influenced by terrestrial environments, e.g., available flight altitudes, ground obstacles, aircraft noise, and weather conditions \cite{Bauranov2021,Tang2022}.  Generally, these factors can be categorized into safety, environmental, and social concerns. Among them, safety is the most paramount factor in low-altitude operations with meteorological conditions and ground obstacles being particularly critical. For instance, adverse weather such as strong winds/rain, and low temperatures can reduce aircraft endurance, impair battery life, and increase collision risks. Although weather forecasting and pre-planned trajectories can mitigate some unfavorable impacts, real-time predictions remain challenging. Moreover, environmental factors like noise and visual pollution also play a significant role in public acceptance of LAWNs, especially in urban areas. The work of \cite{Schaffer2021} showed that drone noise has become a key concern, as the public is more sensitive to drone noise than to vehicle noise. In parallel, visual pollution is another significant issue when dense drones fly overhead, affecting public emotions. 
On top of this, social factors, particularly in terms of privacy and equality preservation, influence public acceptance as well. UAVs' ability to capture personal data during operations raises privacy concerns, and low-altitude services are often perceived as a privilege for high-income groups, leading to resistance from lower-income groups \cite{Sunil2015}. To address these challenges, existing research has structured LAA into five primary types based on various principles. i.e., pipeline airspace, corridor airspace, layered airspace, blocky airspace, and free airspace \cite{juntong2025low}. Each type presents unique design challenges and advantages, and various studies have explored their feasibility, limitations, and applications.
\begin{figure*}
    \centering
    \includegraphics[width=\linewidth]{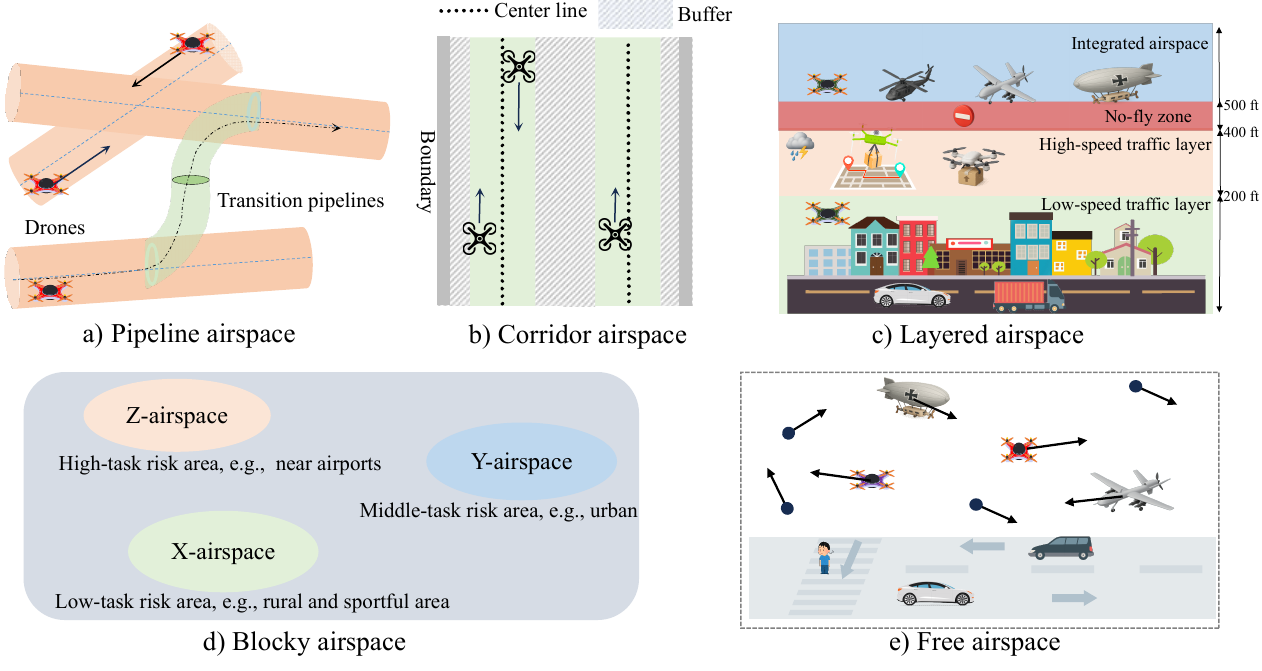}
    \caption{Several representative LAA structures. }
    \label{figLAA}
\end{figure*}

\textit{1) Pipeline Airspace:}
As shown in Fig.  \ref{figLAA}(a), pipeline airspace is the most commonly used form of structured airspace, typically characterized by the establishment of flight path structures or virtual three-dimensional zones within the LAA, akin to road networks on the ground. The pipeline does not necessarily adhere to a fixed geometric shape, and its width is generally designed to accommodate only a single aircraft at a time. Aircraft within the pipeline operate in a unidirectional flight mode and can adjust their altitude, heading, and other parameters at designated intersection points. Notably, pipeline airspace offers a low degree of operational freedom but ensures high safety levels. For instance, the Metropolis project in Europe, which serves as a representative model of pipeline airspace, has designed independent pipelines for both uncrewed and manned aircraft \cite{metropolis2014}. The pipeline airspace consists of several key components: connector pipelines linking the ground to the lowest pipeline layer for takeoff and landing; horizontal pipelines connecting nodes within the same pipeline layer; and transition pipelines bridging different pipeline layers. When an aircraft passes through a marked node, it occupies that node for a specified period, with only one aircraft permitted to occupy the node at any given time \cite{Sunil2015}. Each node maintains an interval list to track the expected occupancy duration of other aircraft. Airspace resources are allocated based on the sequence in which aircraft enter the node, and new aircraft are only allowed to pass through once the node's time slot becomes available. It is important to note that the number of marked nodes in each pipeline layer increases as the altitude decreases for safety.


\textit{2) Corridor Airspace:}
 Compared to pipeline airspace, corridor airspace offers a broader coverage with fewer establishment requirements. As observed in Fig.  \ref{figLAA}(b), it is typically organized in a regular geometric shape, characterized by two key components in terms of a centerline and a width for meeting certain performance criteria. The corridor airspace allows multiple aircraft to fly in parallel following a bidirectional flight pattern. Moreover, corridors can be designed as single or multi-layered airspace to meet traffic flow requirements to increase LAWN capacity. Corridor airspace offers significant advantages such as fewer flight restrictions and greater resilience, where the horizontal and vertical separations among drones are ensured, allowing for maximum flight freedom under safe operational conditions \cite{airbus2018}. To further unlock the potential of corridor airspace, new configurations are investigated to accommodate higher traffic demands. Embraer-X, a Brazilian aerospace company, has proposed dividing LAA into corridors, road segments, and boundary points. In particular, the corridors are dynamically determined by GPS waypoints to fulfill various air traffic flow requirements and improve system efficiency \cite{ embraerx2019}. The authors in \cite{muna2021} introduced multi-layer air corridors, divided into top, middle, and bottom layers, in which the top and bottom layers are designated for drone flying purposes along north-south and east-west directions, respectively. While the middle layer serves as a transitional area for aircraft to temporarily wait and change directions, and thus effectively reducing the collision risk.

\textit{3) Layered Airspace:}
 Layered airspace refers to the division of LAA into several separate layers, each with the same or varying vertical intervals. Since layered airspace is primarily divided in the vertical direction, the operational rules of aircraft in each layer can be effectively controlled, and aircraft operating at different altitudes will not interfere with each other. It is noteworthy that this structured design enhances airspace safety and lowers the technical requirements at the expense of freedom of airspace utilization degradation. In response to the drone market requirements, Amazon proposed a layered LAA design for lightweight drones operating below 500 ft \cite{amazon2015}. In particular, the framework consists of a low-speed traffic layer (below 200 ft) for non-transport missions or UAVs with limited onboard capabilities, a high-speed transport layer (200-400 ft) for faster UAVs capable of autonomous navigation and conflict avoidance in complex environments, and a no-fly zone (400-500 ft) restricted to emergency use, as depicted in Fig.  \ref{figLAA}(c).

\textit{4) Blocky Airspace:}
Block-shaped airspace refers to the division of LAA into horizontal blocks to align with urban structures or specific functional requirements. These blocks differ in infrastructure technology and service provider configurations. Compared to corridor and layered airspace, block-shaped airspace offers greater flexibility, with fewer operational restrictions, allowing more efficient use of airspace. However, block-shaped airspace does not inherently provide aircraft separation,  incurring significant challenges to effective airspace management and conflict avoidance. To achieve this, the SESAR CORUS project \cite{corus2019} introduced X, Y, and Z block-shaped airspaces by jointly considering environmental/social factors and the offered service priority, as shown in Fig. \ref{figLAA}(d). X airspace lies in the lowest service level with aircraft with low task risk, e.g., rural and sportful areas. By contrast, Y airspace is a higher level with strategic conflict avoidance services like those in urban areas. Z airspace is on the top, which supports larger traffic flows and mitigates extensive risk, e.g., the city center and near airports.
\begin{table*}[t]
{
\caption{{Comparison of Representative Airspace Types}}
\centering
\footnotesize
\setlength{\tabcolsep}{4pt}
\renewcommand{\arraystretch}{1.12}
\begin{tabular}{p{2.3cm} p{2.7cm} p{1.7cm} p{1.7cm} p{7.5cm}}
\toprule
\textbf{Airspace Type}     & \textbf{Structure}                                          & \textbf{Capacity}                                       & \textbf{Safety}                                       & \textbf{Advantages \& Disadvantages}                                                       \\ \midrule
\textbf{Pipeline Airspace} & Virtual 3D zones                     & Low                                                  & High                                                & High control, low management cost; Limited flexibility, prone to congestion                    \\ 
\textbf{Corridor Airspace} & Fixed parallel paths                         & Moderate                                             & Moderate                                           & Fewer flight restrictions; Requires extensive planning and management at intersections                  \\ 
\textbf{Layered Airspace}  & Vertical layers                                  & Relatively high                                              & Moderate                                           & Reduces conflict, simple control; Reduces airspace efficiency, less freedom of use                           \\ 
\textbf{Blocky Airspace}   & Horizontal blocks                               & High                                                 & Relatively low                                       &  Accommodate urban development; Complex management, no inherent separation of aircraft                      \\ 
\textbf{Free Airspace}     & Fully flexible                         & Very high                                             & Low                                                & Minimal restrictions, high capacity; Requires advanced collision avoidance, difficult to manage in dense areas    \\ \bottomrule
\end{tabular}
\label{tab:lawn_airspace_comparison}}   
\end{table*}

\textit{5) Free Airspace:}
Different from the above structured LAA, free airspace is a non-structured framework that does not have fixed flight paths or divisions in either the horizontal or vertical direction. In the early 21st century, the National Aerospace Laboratory (NLR) of the Netherlands proposed free flight in the civil aviation domain \cite{hoekstra2001}. The core objective is to enable aircraft to autonomously separate and avoid conflicts through the system, allowing enhanced freedom to select preferred flight paths. Based on this design, the Metropolis project \cite{metropolis2014} proposed free LAA illustrated in Fig. \ref{figLAA}(e). This mechanism only specifies the allowed flight altitude and takes responsibility for managing airspace capacity and setting airspace boundaries. The collision avoidance and aircraft separation are fully addressed by the drones themselves. Due to its unstructured nature, the capacity of free LAA is significantly enhanced. More importantly, the drones only face minimal flight restrictions, and thus significantly shorten their flight distances to the destinations. However, the successful application of free airspace requires a well-developed collision avoidance mechanism. As the missions become complicated, ensuring real-time collision avoidance in dense LAWN deployments remains an open issue. For ease of exposition, we summarize the main features of the aforementioned airspace types in TABLE \ref{tab:lawn_airspace_comparison}, which provides a concise comparison by focusing on their structure, capacity, safety considerations, and a balanced overview of their advantages and disadvantages.

\subsection{Low-Altitude ATM}
\begin{figure}
    \centering
    \includegraphics[width=\linewidth]{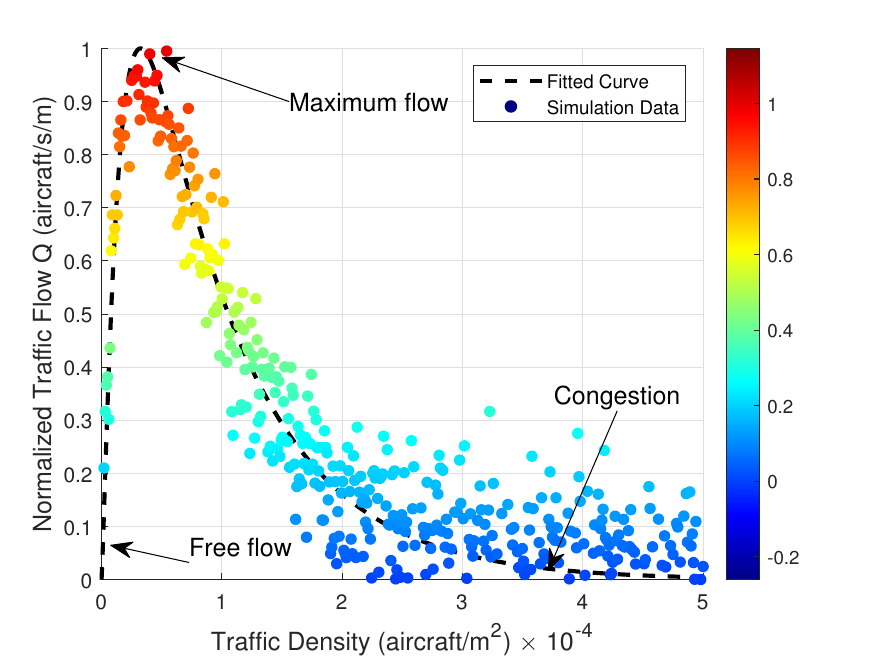}
    \caption{MFD results for the relationship between normalized flow and aircraft density.}
    \label{fig:placeholder}
\end{figure}

{
Apart from effective LAA structuring, the management of LAA has become a critical issue with the rise of UAM. To address the challenges associated with increasing traffic densities, macroscopic fundamental diagrams (MFDs), originally developed for ground traffic management, have been adapted for LAA \cite{Rastgoftar2024}. An MFD characterizes the aggregate relationship among traffic flow, density, and speed at the network level and thereby provides a compact description of the operational state of the airspace. As shown in Fig.~\ref{fig:placeholder}, the simulation data and the fitted curve jointly reveal a characteristic three-regime behavior. In the free-flow regime, at low densities, the normalized traffic flow increases almost linearly with density. As density approaches a critical value, the flow reaches a maximum, which indicates the practical capacity of the airspace. Beyond this critical density, the system enters a congested regime, where additional aircraft lead to a reduction in flow and the emergence of unstable operating conditions. This relationship demonstrates that MFDs can capture capacity limits and congestion onset in LAA, and thus provide a useful tool for designing density thresholds, scheduling policies, and flow control strategies for LAWNs.}
The work of \cite{Bulusu2018} presented a throughput-based capacity metric to evaluate the impact of increasing air traffic on LAA. The proposed method simulated traffic and analyzed the effects of different conflict resolution methods, where the findings suggest that throughput decreases as traffic density rises, even while maintaining safety, offering a new way to assess conflict detection strategies.  The authors of \cite{Li2022} addressed the challenges faced by the national airspace system in optimizing traffic flow, in which various strategies for managing air traffic are reviewed. The authors emphasized the role of collaborative decision-making (CDM) and automation to improve capacity and reduce congestion in high-density environments. Furthermore, the study of \cite{Work2009} introduced a new approach to air traffic flow optimization using convex formulations of partial differential equations (PDEs). The authors modeled air traffic flow in an Eulerian network and demonstrated how to convert these models into linear and quadratic optimization problems, providing globally optimal solutions for large-scale air traffic systems.

A machine learning-based decision support system for ATM was further developed by integrating explainable AI (XAI) in \cite{aerospace8080224}. By employing learning-based algorithms to predict risks, this approach can assist controllers in managing air traffic effectively. The incorporation
of XAI ensures that the reasoning behind AI decisions is transparent and understandable to human operators, thereby increasing trust in automated systems. Furthermore, the research explored the system’s application in both controlled and uncontrolled airspace, illustrating its potential to enhance safety and operational efficiency in low-altitude ATM. To achieve efficient ATM in real-world environments, significant reforms are being made in the strategic planning and development of ATM systems under the single European sky (SES) initiative \cite{REZO20235}, motivated by the need to establish unified strategic objectives to improve LAWNs' capacity, safety, and environmental sustainability. A comprehensive review of the ATM planning framework in Europe has revealed critical gaps, particularly regarding the alignment of performance targets with local operational needs. The findings suggest that current strategic goals for capacity are overly broad and fail to address the specific requirements of individual regions and airspaces. To sum up, although the low-altitude ATM development is progressing, it remains in the early stages to support practical implementation.

{Moreover, the deployment of LAWNs necessitates a strategic alignment with both international frameworks and domestic regulations to ensure safety, interoperability, and compliance. A comprehensive review of representative standards relevant to LAWNs is provided to demonstrate how such systems can seamlessly integrate into global and national UAV ecosystems while adhering to regulatory requirements. The ICAO has developed the ATM framework, which outlines essential capabilities for managing UAVs within LAA \cite{icao_utm_framework}. This framework emphasizes the need for global harmonization in airspace management and the integration of UAVs into existing ATM systems. Key capabilities within the ATM system include flight authorization, real-time information sharing, conflict management, and automated traffic de-confliction, all of which must be supported by LAWNs to ensure safe and coordinated UAV operations. By doing so, LAWNs can ensure the safe management of UAVs, complementing the existing aviation infrastructure.}

\begin{figure}
    \centering
    \includegraphics[width=\linewidth]{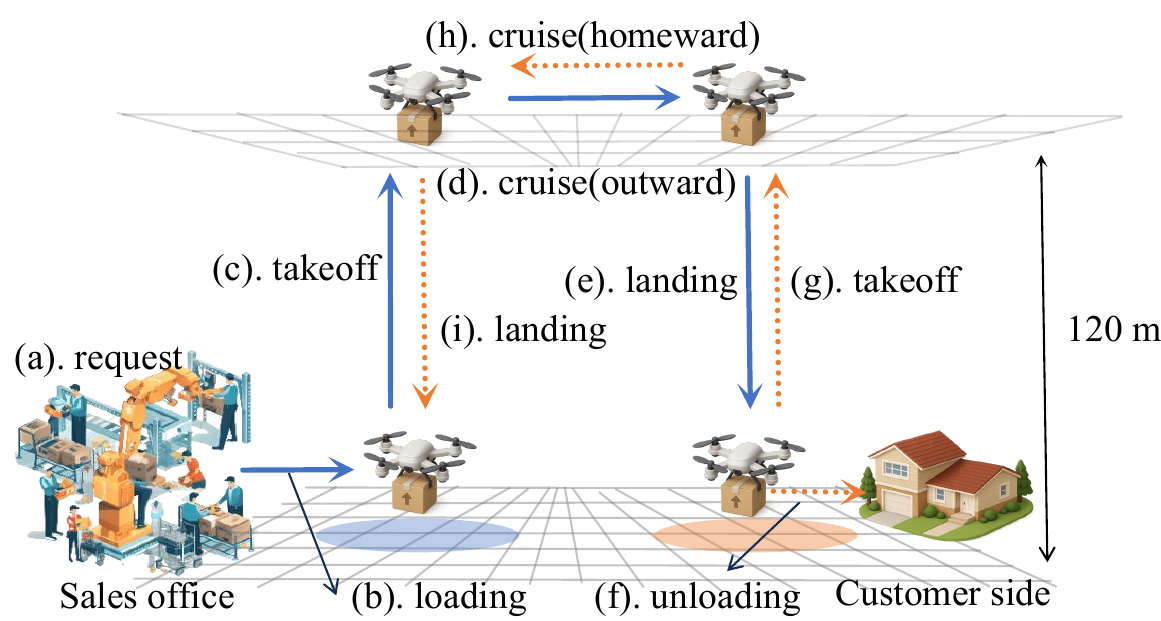}
    \caption{The typical UAV delivery workflow of Ref. \cite{oosedo2021unmanned}.}
    \label{fidig}
\end{figure}

{In the United States, the Federal Aviation Administration (FAA) introduced its ATM concept of operations \cite{FAA2022}, envisioning a network of UAV service suppliers that provide services such as digital flight planning, airspace management, and UAV tracking. The FAA’s framework operates through cloud-to-cloud data exchanges, eliminating the need for traditional voice communication with air traffic control. A key feature of the FAA system is Remote ID, which mandates UAVs to broadcast unique identification and location data to enable real-time tracking and compliance with airspace restrictions. LAWNs must integrate with these services, ensuring that UAVs are monitored, controlled, and managed in accordance with FAA regulations. To support studies on future parcel delivery by UAV, the Japan Aerospace Exploration Agency (JAEA) has introduced a dedicated simulation platform for the 2030 timeframe, termed the scalable simulator for knowledge of low-altitude environment (SKALE). This framework emulates BVLoS operations above densely populated urban regions. As illustrated in Fig. \ref{fidig}, a representative delivery scenario is configured with a nominal cruise altitude of 120 m. Analysis based on SKALE indicates that current conflict detection and resolution strategies are not yet sufficient for large-scale, highly automated traffic. In particular, the appropriate setting and modelling of separation minima and flight speeds remain open research problems that require further refinement.}
{Similarly, the Civil Aviation Administration of China (CAAC) has implemented regulations that follow a risk-based approach to managing LAA \cite{caac_low_altitude_service_system_plan_2018}. UAVs are categorized based on their weight and operational risk level, with all UAVs, except the smallest, required to be equipped with unique electronic identification and registered under a real-name system.  In addition, LAWNs are required to integrate geofencing capabilities and enforce remote flight restrictions according to the operational limitations set by regulatory authorities. The policies within China are actively driving the LAE development, generating significant demand for high-capacity, scalable LAWNs that can support dense UAV applications in terms of drone delivery, eVTOL air taxis, and remote sensing. To meet these requirements, LAWNs must provide ultra-reliable communication and low-latency connectivity to ensure seamless integration and compliance with regulatory standards in an expanding LAA.
Toward this end, the 3rd Generation Partnership Project (3GPP) has released several technical reports (TRs) addressing LAWN communications \cite{3gpp_tr_36777}. Notably, TR 36.777 explores the UAV connectivity by leveraging LTE, highlighting key challenges such as interference and mobility. Furthermore, TR 38.811 and TR 38.821 focus on the integration of 5G new radio (NR) within non-terrestrial networks (NTN), which includes satellites and HAPs.
} 

\subsection{Emerging Path Planning Technology}

The preceding discussion on low-altitude ATM focuses on macroscopic aspects such as capacity, separation policy, and regulatory coordination. In practice, these system-level concepts are ultimately realized through the individual trajectories flown by each aerial vehicle. Therefore, path planning forms the microscopic counterpart of ATM and directly determines whether a low-altitude airspace can accommodate dense operations without compromising safety or efficiency. A well-designed trajectory must remain compatible with corridor structures, restricted zones, and dynamic flow constraints, while obeying platform limits on energy, payload, and maneuverability. In the context of LAWNs, where numerous heterogeneous UAVs operate in close proximity, deficiencies in path design manifest as conflict hot spots, unnecessary detours, and excessive delay. From a design perspective, most emerging path planning technologies are driven by two primary objectives, namely, trajectory safety and trajectory efficiency \cite{aggarwal2020path}. Specifically, safety-oriented planning emphasizes collision and obstacle avoidance, while compliance with no-fly regions and separation minima. On the other hand, efficiency-oriented planning seeks to reduce mission cost, for example, by shortening flight time and lowering energy consumption through multiobjective criteria. In many recent studies, safety is enforced through hard constraints, whereas efficiency guides the selection of a preferred solution within the feasible set.

For safety-oriented path planning, configuration-based methods remain a central tool, in which the state of a drone, including its position, heading, altitude, and possibly velocity, is embedded into an abstract configuration space in which each point represents a feasible geometric configuration of the vehicle. Physical obstacles, terrain, and no-fly regions are mapped to forbidden subsets of this space, and path planning is formulated as the problem of finding a continuous curve in the free subset that connects the initial and target configurations while obeying kinematic and dynamic constraints. This geometric abstraction allows explicit treatment of vehicle size and maneuverability and provides clear guarantees on collision avoidance once a path in the free configuration space has been obtained. For instance, the work in \cite{67} refined the classical artificial potential field method by introducing a distance-dependent repulsive function, which increases clearance from obstacles and alleviates local minima in dense environments. The authors decomposed the repulsive effect into separate speed and heading components and combined it with a collision prediction mechanism, which yielded smoother trajectories while preserving strong safety margins.  The work in \cite{69} employed an improved jump point search model for multirotor UAVs and demonstrated that the resulting paths were both shorter and more maneuverable than those obtained with standard A* search in static urban scenes. To overcome the limitations of two-dimensional grids, the authors in \cite{70} extended jump point search to a three-dimensional representation and proposed modified parent node expansion and pruning strategies, which significantly increased spatial flexibility and computational efficiency, in which the risk, acoustic impact, and logistics cost were incorporated into an A*-based planner for urban delivery missions and showed that trajectories can be shaped to satisfy safety and environmental constraints while retaining practical runtime.

In the context of multiple aircraft, the cooperative and non-cooperative methods can explicitly capture interactions and support online conflict resolution in dynamic environments. The authors in \cite{72} formulated dynamic collision avoidance in high-density urban low-altitude airspace as a Markov decision process and integrated the resulting algorithm into an operational system, achieving real-time conflict management. Moreover, the work in \cite{73} proposed a genetic algorithm-based route replanning method for scenarios with unknown obstacles, where the appearance of new hazards triggers a global search for alternative paths that maintain feasibility and separation. The study in \cite{74} introduced a two-layer reinforcement learning model, in which a local layer performs short-horizon obstacle avoidance and a global layer selects long-horizon routes, enabling simultaneous handling of static and dynamic obstacles with improved path continuity. 

By contrast, efficiency-oriented path planning methods build on this safety foundation and place additional emphasis on operating cost and computational complexity. For instance, the work in \cite{77} modeled UAV path planning under partial observability as a partially observable Markov decision process and sought trajectories that minimize cumulative operating cost while accounting for nonlinear dynamics and measurement uncertainty. This formulation provides a principled way to balance performance and robustness when the vehicle has only partial knowledge of its environment. Along this line, the work in \cite{cui2014improved} observed that traditional UAV path planners often ignore mission-specific constraints such as terminal heading and proposed a bidirectional ant-colony-optimization scheme on a grid-based workspace to address this limitation. Ant colonies are initialized near the start and goal configurations and expand simultaneously along predetermined directions, after which a dedicated path-selection mechanism combines and filters candidate routes. Improved pheromone-update rules and successor-selection strategies were introduced to accelerate convergence and reduce the risk of local optima. From a cooperative perspective, the authors in \cite{belkadi2019design} developed a distributed controller for a fleet of UAVs under a multi-agent-systems framework, where online path planning is formulated as a particle-swarm-optimization (PSO) problem solved independently on each vehicle. By minimizing a suitably designed cost function, each local PSO instance determines a trajectory that contributes to formation keeping, target tracking, and collision avoidance, and experiments with fixed and moving targets, external disturbances, and agent loss confirmed that the distributed strategy can maintain coherent fleet behavior and safety without centralized coordination.
Overall, emerging path planning technologies complement low-altitude ATM by bridging system-level policies with executable trajectories. Safety-oriented configuration-space methods and cooperative decision frameworks provide reliable guarantees in structured and moderately dynamic airspace, whereas efficiency-oriented approaches enhance scalability and performance in large-scale, high-density LAWNs. The combined use of these approaches is expected to be essential for supporting dense, task-driven operations while maintaining both operational safety and mission effectiveness.

\section{Open Challenges}\label{sec8}

\subsection{Digital Twin-Enabled LAWNs}
The integration of digital twin (DT) technologies with LAWNs presents a promising approach for real-time modeling, monitoring, and predictive control of UAV swarms and aerial infrastructures. By constructing virtual replicas of UAVs, communication links, and environmental conditions, DTs enable proactive network management, fault prediction, and efficient trajectory optimization. However, realizing DT-enabled LAWNs faces several challenges, including the need for high-fidelity modeling of dynamic airspace, synchronization between virtual and physical entities, and the significant computational and communication overhead required for maintaining real-time updates. 

\subsection{Lightweight AI-Enhanced LAWNs}
Although integrating large AI models into LAWNs can enable intelligent decision-making, autonomous mission planning, and semantically aware resource allocation on UAVs and edge nodes, practical deployment at the network edge is severely constrained by limited computation, energy budgets, and model size. These constraints make straightforward deployment of current large models infeasible on mobile aerial platforms. Consequently, the design of lightweight and energy-efficient LLM architectures, together with distributed inference frameworks tailored to UAV and edge hardware, remains a central open problem for realizing next-generation intelligent LAWNs.

\subsection{Standardizations for LAWN Deployments}
Despite the significant potential of LAWNs in enabling LAE applications, their large-scale adoption is hindered by the lack of standardized frameworks and regulatory guidelines. Key challenges include defining airspace usage protocols, ensuring interoperability among heterogeneous UAV platforms, allocating spectrum for aerial communications, and ensuring compliance with safety and privacy regulations. Standardization efforts must strike a balance between promoting innovation and ensuring regulatory compliance, enabling the reliable, scalable, and secure deployment of LAWNs. Furthermore, the global harmonization of technical, operational, and legal requirements across various regions remains a critical issue that must be addressed to fully unlock the commercial potential of LAWNs.
{
\subsection{3D Interference management}
As UAVs operate in 3D airspace, managing interference in LAWNs becomes increasingly complex due to the dynamic nature of both UAV movements and communication channels. Effective 3D interference management is critical to ensure reliable communication, reduce latency, and optimize network capacity. This involves not only minimizing interference between UAVs and ground stations but also managing interactions between UAVs at various altitudes and those operating in adjacent airspaces. Key challenges in 3D interference management include real-time spectrum allocation, adaptive beamforming, and interference mitigation techniques that account for the unique mobility patterns of UAVs.
\subsection{Multi-Domain Traffic Management}
Multi-domain traffic management (MDTM) seeks to harmonize advanced low-altitude airspace management with conventional air traffic management and emerging space traffic management. LAWNs must share consistent rules, intent information, and safety constraints with vehicles operating across different altitudes and time scales, which is complicated by heterogeneous standards and legacy systems. Therefore, designing unified interfaces and cross-domain conflict resolution mechanisms remains intractable for embedding LAWNs into an integrated air-and-space traffic ecosystem.
\subsection{Low-Altitude 4D Trajectory Optimization}
Low-altitude four-dimensional (4D) trajectory optimization aims to jointly design three-dimensional flight paths and time profiles so that UAVs satisfy mission deadlines, separation constraints, and airspace capacity limits in a coordinated manner. In dense LAWNs, this problem becomes highly coupled across vehicles and strongly affected by uncertainties in demand, weather, and communication quality. Developing scalable optimization and learning-based methods that can generate conflict-free, energy-efficient 4D trajectories in real time, while remaining compatible with existing ATM and path-planning frameworks, remains a key open research challenge.
\subsection{Low-Altitude Traffic Flow Prediction}
Low-altitude traffic flow prediction aims to forecast UAV demand and spatiotemporal flow patterns across corridors and sectors, providing critical inputs for capacity planning, congestion mitigation, and conflict management in LAWNs. Compared with traditional air traffic, low-altitude flows exhibit strong nonstationarity, tight coupling with urban activities and weather, and sparse or heterogeneous data sources. Developing reliable data-driven and model-based prediction frameworks that can fuse multi-source information, quantify uncertainty, and operate in real time remains a key open challenge for scalable low-altitude traffic management.}

\section{Conclusion}\label{sec9}

This paper provided a comprehensive survey of LAWNs, focusing on their functional designs and LAA management. We began by comparing LAWNs with traditional aerial communication systems, highlighting their ability to integrate communication, sensing, computation, and control. We then established the foundational principles for LAWN architectures and channel modeling, followed by a review of recent advancements in performance metrics, functional designs, security/privacy preservation, and AI-driven approaches. Finally, we summarized methodologies for LAA structuring, traffic management, and path planning, essential for ensuring scalable and safe LAWN operations.

\section*{Acknowledgment}
This work is supported in part by National Natural Science Foundation of China under Grant 62501600 and 62471208, in part by Guangdong Provincial Natural Science Foundation under Grant 2024A151510098, in part by Shenzhen Science and Technology Program under Grant JCYJ20240813094627037. 

\bibliography{myref}{}

\bibliographystyle{IEEEtran}

\end{document}